\documentclass[apj]{emulateapj}
\usepackage{amsmath}
\usepackage{epstopdf}
\usepackage{dpfloat}

\shorttitle{The Shearing Spiral Arms of NGC 1365}
\shortauthors{Speights, Westpfahl}

\begin{document}

\title{The Shearing H{\hskip 1.5pt \footnotesize  I} Spiral Pattern of NGC 1365}

\author{Jason C. Speights {\scriptsize AND} David J. Westpfahl}
\affil{Department of Physics, New Mexico Institute of Mining and Technology, Socorro, NM 87801; jspeight@nmt.edu}

\begin{abstract}

The Tremaine-Weinberg equations are solved for a pattern speed that is allowed to vary with radius.  The solution method transforms an integral equation for the pattern speed to a least squares problem with well established procedures for statistical analysis.  The method applied to the H{\hskip 1.5pt \footnotesize  I} spiral pattern of the barred, grand-design galaxy NGC 1365 produced convincing evidence for a radial dependence in the pattern speed.  The pattern speed behaves approximately as 1/$r$, and is very similar to the material speed.  There are no clear indications of corotation or Lindblad resonances.  Tests show that the results are not selection biased, and that the method is not measuring the material speed.  Other methods of solving the Tremaine-Weinberg equations for shearing patterns were found to produce results in agreement with those obtained using the current method.  Previous estimates that relied on the assumptions of the density-wave interpretation of spiral structure are inconsistent with the results obtained using the current method.  The results are consistent with spiral structure theories that allow for shearing patterns, and contradict fundamental assumptions in the density-wave interpretation that are often used for finding spiral arm pattern speeds.  The spiral pattern is winding on a characteristic timescale of $\sim$ 500 Myrs.
 
\end{abstract}


\section{INTRODUCTION}%1

The purpose of this paper is to develop and apply a new method of solving the Tremaine-Weinberg equations (Tremaine \& Weinberg 1984, hereafter TW84) for a pattern speed, $\Omega_p$, that is allowed to vary with radius.  The pattern speed is an important astrophysical parameter for understanding the role of spiral and bar structures in the formation, dynamics, and evolution of a galaxy.  Its radial behavior is not well understood.  

The method is capable of checking fundamental assumptions in the density-wave interpretation of spiral structure that are often used for measuring $\Omega_p$.   This interpretation assumes that $\Omega_p$  is constant (i.e., independent of radius ), and that resonant radii are identifiable on a rotation curve of the galaxy (see Bertin \& Lin 1996; Bertin 2000).  Resonant radii are assumed to occur when $\Omega_p$ = $\Omega$, the material speed, and when $\Omega_p$ $=$ $\Omega$ $\pm$ $\kappa$/$m$, where $\kappa$ is the epicycle frequency of a stellar orbit and $m$ is the mode of the wave corresponding to the number of arms \cite{bl38}.  These resonances are commonly referred to as corotation and Lindblad resonances respectively, and are identified by interpreting photometric and kinematic features of a galaxy.  The assumption that $\Omega_p$ is constant in the spiral arms has yet to be confirmed by direct observation.

An alternative way of measuring $\Omega_p$ is to solve the model-independent TW84 equations.  These are integrated forms of the continuity equation that relate $\Omega_p$ to the brightness of a pattern tracer and the observable line-of-sight velocity.  When $\Omega_p$ is allowed to vary with radius, integrating along paths that are parallel to the kinematic major axis of the galaxy produces an integral equation for $\Omega_p$,
%e1
\begin{eqnarray}
\int_{-\infty}^{+\infty}{\Omega_p(r)\Bigl (x{\partial \over \partial y}I(x,y) - y{\partial  \over \partial x}I(x,y)\Bigr)}\,dx& \nonumber \\ =  \int_{-\infty}^{+\infty}{\partial \over \partial y}I(x,y) \,v_y(x,y)\,&dx,
\end{eqnarray}
where $I$ is the specific intensity of a pattern tracer and $v_y$ is the line-of-sight velocity times the cosecant of the disk inclination.  An illustration showing how a galaxy's disk is oriented in the coordinate system of Equation (1) is shown in Figure 1.  In this coordinate system, $y$ is measured in the plane of the disk.  Equation (1) is similar to Equation (4) in TW84 with the exception that $\Omega_p$ is allowed to vary with radius.  Although allowing $\Omega_p$ to vary with radius may violate assumption 2. in TW84, it can be shown that the derivation by TW84 is generalizable to a pattern speed that varies with radius (for a proof see Engstr\"{o}m 1994, hereafter E94).  The steps involved in deriving Equation (1) for  H{\hskip 1.5pt \footnotesize  I} are shown in Section 2.2.   

The new solution method uses least squares to solve Equation (1) for simple functional forms of $\Omega_p$.  The two forms solved for in this paper are a polynomial in $r$ = $\sqrt{x^2+y^2}$,
%e2
\begin{equation}
\Omega_p(r) \, =  \sum_{i=0}^{n}\alpha_i r^{i},
\end{equation} 
and a polynomial in 1/$r$,
%e3
\begin{equation}
\Omega_p(r) \, =  \sum_{i=0}^{n}{\alpha_i \over r^{i}},
\end{equation} 
where $\alpha_i$ are unknown constants to be determined in the solution.  Both equations can represent a constant $\Omega_p$ when $n$ = 0.  Although they are simple in their form, Equations (2) and (3)  are adequate for investigating whether $\Omega_p$ has a radial dependence.  If evidence for such a dependence is found, then the pattern is shearing, and the solution obtained when $n$ = 0 is the mean pattern speed.  

%f1
\begin{figure}
\centering
\includegraphics[width=0.47\textwidth]{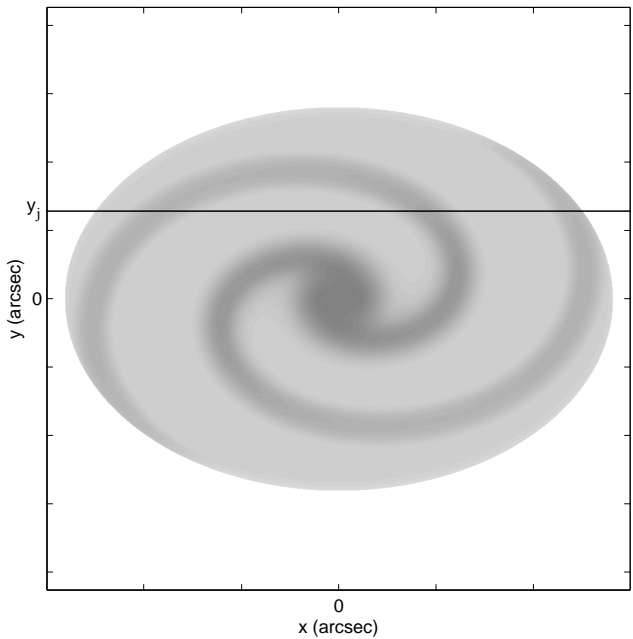} 
\caption{Illustration showing how a galaxy disk is oriented in the coordinate system of Equation (1).  The origin of the coordinate system is at the galaxy's kinematic center.  The kinematic major axis of the disk is along the $y$ = $0$ axis.  An example of an integration path at $y$ = $y_j$ is shown as a solid line across the disk.} 
\end{figure}

To find the least squares solution, first combine Equations (1) and (2) to form a sum on the left hand side of Equation (1), 
%e4
\begin{eqnarray}
\sum_{i=0}^{n}\alpha_i\int_{-\infty}^{+\infty}{r^{i}\,\Big\{x{\partial \over \partial y}I(x,y) - y{\partial \over \partial x}I(x,y)\Big\}\,dx}& \nonumber \\ =  \int_{-\infty}^{+\infty}{\partial\over \partial y}I(x,y)v_y(x,y)\,d&x,
\end{eqnarray}
and similarly for Equation (3),
%e5
\begin{eqnarray}
\sum_{i=0}^{n}\alpha_i\int_{-\infty}^{+\infty}{{1 \over r^{i}}\,\Big\{x{\partial \over \partial y}I(x,y) - y{\partial \over \partial x}I(x,y)\Big\}\,dx}& \nonumber \\  =  \int_{-\infty}^{+\infty}{\partial\over \partial y}I(x,y)v_y(x,y)\,d&x.
\end{eqnarray}
What remains inside the integrals are observable quantities that can be calculated at different $y$ = $y_j$ distances from the kinematic major axis (see Figure 1).  For several independent calculations let,
%e6
\begin{equation}
\mbox{G}_{j,i}=\int_{-\infty}^{+\infty}{r^{i}\Big\{x{\partial \over \partial y}I(x,y) - y{\partial \over \partial x}I(x,y)\Big\}}\bigg|_{y_j}dx, 
\end{equation}
for Equation (4), and similarly,
%e7
\begin{equation}
\mbox{G}_{j,i}=\int_{-\infty}^{+\infty}{{1 \over r^{i}}\Big\{{\partial \over \partial y}I(x,y) - y{\partial \over \partial x}I(x,y)\Big\}}\bigg|_{y_j}dx, 
\end{equation}
for Equation (5), where that matrix indexing is used for $j$ rows and $i$ columns.  Also let, 
%e8
\begin{equation}
\mbox{d}_{j}= \int_{-\infty}^{+\infty}{\partial\over \partial y}I(x,y)v_y(x,y)\bigg|_{y_j}dx,
\end{equation}
for the right hand sides of Equations (4) and (5).  In Equations (6) and (7) the $i$th column corresponds to the power of $r$ in $\Omega_p$.  In Equations (6), (7), and (8) the $j$th row corresponds to a calculation at $y_j$.  Using these matrices, Equations (4) and (5) in matrix form are,
%e9
\begin{equation}
{\bf G \boldsymbol{\alpha} = d},
\end{equation}
where that $\boldsymbol \alpha$ $=$ $\alpha_i$ are the unknown coefficients.  Multiplying both sides of Equation (9) by the transpose of {\bf G} produces the normal equations, 
%e10
\begin{equation}
{ \bf G}^{T}{\bf G}{\boldsymbol \alpha} = {\bf G}^{T}{\bf  d},
\end{equation}
for the least squares problem.  The maximum likelihood solution for the best-fitting coefficients is obtained by multiplying both sides of Equation (10) by ({\bf G}$^T${\bf G})$^{-1}$.  Standard statistical tools can then be used to analyze the results and determine whether a constant $\Omega_p$, or one that varies with $r$, provides a better solution.  

The rest of this paper is dedicated to applying the solution method to the spiral pattern of NGC 1365 and analyzing the results.  In Section 2 is a description of NGC 1365 and the details of applying the method.  In Section 3 is an explanation of the statistics used for estimating the uncertainties and analyzing the results.  In Section 4 are the results and their statistical analysis.  In Section 5 the method is applied to different regions of NGC 1365 to rule out selection bias in the results.  For all of the regions the method was applied to, the results are very similar to $\Omega$.  In Section 6, therefore, the method is tested using galaxies without spiral patterns to see if their results are also similar to $\Omega$.  In Section 7 other solution methods are applied to NGC 1365 and the results are compared to those obtained using the current method.  In Section 8 the results are compared with previous estimates of $\Omega_p$.  In Section 9 the relevance of the results to theories of spiral structure is discussed, and in Section 10 is a summary.
	
\section{APPLICATION TO NGC 1365}%2

\subsection{Description of NGC 1365}%2.1

NGC 1365 is a grand-design spiral galaxy of type SB(s)b\cite{v91}.  An optical image of NGC 1365 is provided in Figure 2 to assist in its description.  The image is reproduced from the Southern Digitized Sky Survey of the Space Telescope Science Institute\footnote{Original plate material is copyrighted by the Royal Observatory Edinburgh and the Anglo-Australian Observatory. The plates were processed into the present compressed digital form with their permission. The Digitized Sky Survey was produced at the Space Telescope Science Institute under US Government grant NAG W-2166.}.  The highly symmetric, coherent, two-arm pattern of NGC 1365 is ideal for the first application of the solution method.  The disk inclination is small enough to observe the spiral structure, yet large enough to accurately measure $v_y$.  The assumed position, orientation, and systemic velocity, $V_{sys}$, that were used for applying the method are shown in Table 1.  These were adopted from the analysis of the H{\hskip 1.5pt \footnotesize  I} velocity field by J\"{o}rs\"{a}ter \& van Moorsel (1995, hereafter JvM95).  NGC 1365 belongs to the Fornax cluster\cite{bb60}, but has no known close companions.  This archetypical barred spiral galaxy is well studied.  For more information about NGC 1365, see the review by \cite{po99}.

%f2
\begin{figure}[t!]
\centering
\includegraphics[width=0.47\textwidth]{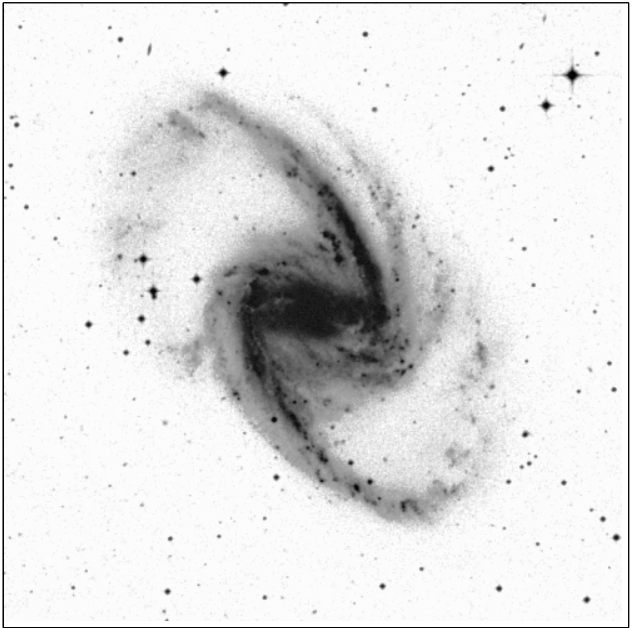} 
\caption{Optical image of NGC 1365 in the blue part of the optical spectrum. The original image is a 60 minute IIIa-J exposure using the UK 48-inch Schmidt Telescope and a 4680 \AA \hskip 1ex GG 395 filter.  North is up and East is to the left.  The receding half of the galaxy is to the bottom right.} 
\end{figure}

\subsection{H{\hskip 1.5pt \footnotesize I} as a Pattern Tracer}%2.2

TW84 derive pattern speed equations from a form of the continuity equation that assumes mass conservation.  This assumption is difficult to satisfy for tracers of spiral patterns; starlight can be obscured by dust, and the ISM is multi-phased.  When using a component of the ISM such as H{\hskip 1.5pt \footnotesize I}, an argument is usually made for dominance in the abundance of the tracer within dynamical (orbital) timescales so that conversion among the different phases of the ISM is negligible.  These arguments were made for H{\hskip 1.5pt \footnotesize I} by Bureau et al. (1999) and Meidt et al. (2009, hereafter M09), and for H$_2$ by Zimmer et al. (2004), Rand \& Wallin (2004), Merrifield et al. (2006, hereafter M06), Meidt et al. (2008b), and Meidt et al. (2009).  For the latter group, a known 
conversion factor from CO surface brightness to H$_2$ abundance is also assumed.  M09 also combined measurements of H{\hskip 1.5pt \footnotesize I} with estimates of H$_2$ in an attempt to account for the total abundance of gas.  The TW84 equations have also been applied to H-$\alpha$ emission by Hernandez et al. (2005), Emsellem et al. (2006), Fathi et al. (2007), Chemin \& Hernandez (2009), Gabbasov et al. (2009), and Fathi et al (2009), the first and the last of which acknowledge that ionized gas neither dominates the ISM, nor persists within dynamical timescales, but claim that it may come close enough. 
%t1
\begin{deluxetable}{llc}[t!]
\tablecaption{Adopted Parameters of NGC 1365}
\tablewidth{0pt}
\startdata
\tableline\tableline\\[-7.5 pt]
Kinematic center R. A. (J2000) && \hskip 3.8pt 03$^{h}$ 33$^{m}$ 36.4$^{s}$ $\pm$ 4$\arcsec$\\
Kinematic center DEC. (J2000) && $-$36$\degr$ 08$\arcmin$ 25.5$\arcsec$ $\pm$ 4$\arcsec$\\
Inclination angle && \hskip 5pt 40$\degr$ $\pm$ 2$\degr$ \\
Position angle && 220$\degr$ $\pm$ 2$\degr$\\
V$_{sys}$  && 1632 $\pm$ 3 km s$^{-1}$
\enddata
\end{deluxetable}

The TW84 equations are derivable, however, from the general form of the continuity equation,
%e11
\begin{eqnarray}
{\partial \over \partial t} I(x,y,t) +{\partial\over  \partial x}I(x,y,t)\,v_x(x,y,t)& \nonumber \\  
+{\partial\over \partial y}I(x,y,t)\,v_y(x,y,t) = S&(x,y,t),
\end{eqnarray}
that includes a source function, $S$, to account for the creation and destruction of material.  This was shown by Westpfahl (1998, hereafter W98), who assumed that the intensity has a solution of the form,  
%e12
\begin{equation}
I(x,y,t) = \tilde{I}(x,y,t) +\int_{0}^{t}S(x,y,t')\,dt',
\end{equation}
where the time dependance of $\tilde{I}$ is due entirely to rotation.  The integral of the source function is differentiable using Leibnitz's rule, 
%e13
\begin{equation}
{\partial\over \partial t}\int_{0}^{t}S(x,y,t')\,dt' = S(x,y,t) + \int_{0}^{t}{\partial \over \partial t} S(x,y,t')\,dt',
\end{equation}
and from TW84 the time derivative of $\tilde{I}$ is,
%e14
\begin{equation}
{\partial \over \partial t} \tilde{I}(x,y,t) = {\Omega_p(r)\Bigl (y{\partial  \over \partial x}I(x,y,t) - x{\partial \over \partial y}I(x,y,t)\Bigr)}.
\end{equation}
Combine Equations (11) through (14), and the result, 
%e15
\begin{eqnarray}
&&\hskip -15pt \Omega_p(r)\Bigl (y{\partial  \over \partial x}I(x,y,t) - x{\partial \over \partial y}I(x,y,t)\Bigr) \nonumber \\ &&\hskip -5.6pt + {\partial\over  \partial x}I(x,y,t)\,v_x(x,y,t) +{\partial\over \partial y}I(x,y,t)\,v_y(x,y,t) \nonumber  \\  &&\hskip 100pt =  - \int_{0}^{t}{\partial  \over \partial t}S(x,y,t')\,dt' \nonumber \\
&&\hskip 100pt =S(x,y,0)-S(x,y,t), 
\end{eqnarray}
is different from that derived in TW84 by the change in the source function on the right-hand side.  For the purpose of this paper, it is assumed that the source function is approximately constant within dynamical timescales.  The result, then, is the same pattern speed equation derived by TW84, but from a form of the continuity equation that does not assume mass conservation.   With the change in the source function set to zero, integration in $x$ removes the velocity term in that direction due to the boundedness of $I$, and produces Equation (1).

Of the spiral pattern tracers that are available, H{\hskip 1.5pt \footnotesize I} has several advantages:

\indent 1.)   Radio interferometers such as the Very Large Array (VLA) of the National Radio Astronomy Observatory (NRAO) are capable of producing high-resolution specific intensity and velocity maps of H{\hskip 1pt \footnotesize  I} from the 21-cm line.  

\indent 2.)  To a good approximation, the 21-cm line is optically thin, unlike, for example, the 2.6-mm line of CO.  Assuming the 21-cm line is optically thin, it is straightforward to measure the total abundance of H{\hskip 1.5pt \footnotesize  I}, which is necessary for applying the continuity equation.  

\indent 3.)  The spiral patterns of galaxies often extend much farther in the 21-cm line than they do in other tracers such as optical starlight, H-$\alpha$ emission, or the 2.6-mm line.  

\indent 4.)  Unlike optical starlight or H-$\alpha$ emission, the 21-cm line is not vulnerable to obscuration and reflection by dust.  

\indent 5.)  Bar patterns are often undetected in H{\hskip 1.5pt \footnotesize I} maps of barred spiral galaxies, allowing for a separate measurement of the spiral arm pattern speed.

\subsection{Data}%2.3

Specific intensity and velocity maps made by JvM95 were used to make maps of $I$ and $v_y$ as described in the Introduction.  They mapped the H{\hskip 1.5pt \footnotesize  I} to high resolution from VLA observations in the BnA, CnB, C, and DnC configurations, amounting to a total of 53.8 hours on source.  The major and minor axis at FWHM of the synthesized beam was 11.55$\arcsec$ $\times$ 6.32$\arcsec$ with a position angle of 5.33$^\circ$.  The RMS noise in a line-free channel was 0.27 mJy beam$^{-1}$.  The channel width was 20.84 km s$^{-1}$.  For more information about the observations and data reduction see JvM95.  

The original maps were modified using the Astronomical Image Processing System (AIPS) developed and maintained by the NRAO.  This included using the AIPS task OGEOM for image translation and rotation, and the  AIPS task MATHS for subtracting $V_{sys}$ from the velocity map and then multiplying it by the cosecant of the disk inclination.  Figure 3 shows the centered and rotated specific intensity map overlaid with contours of constant $v_y$.  The pixel increment spans 2$\arcsec$ in the $x$ direction and 2.61$\arcsec$ in the $y$ direction.  Estimates of the uncertainty per pixel in the maps of $I$ and $v_y$ are explained in Section 3.1.  They are $\sigma_I$ =  0.54 mJy km s$^{-1}$ pixel$^{-1}$, and $\sigma_{v_{y}}$ = 12.86 km s$^{-1}$ pixel$^{-1}$, respectively.
 
\subsection{Implementation}%2/4

The solution method was implemented within MATLAB\footnote{The Mathworks, Version 7.10.0.499 (R2010a), http://www.mathworks.com}.  Calculations of Equations (4) and (5) were restricted to the region $|y|$ $\leqslant$ 400$\arcsec$.  Different $y_j$ were spaced 5 pixels apart, which amounts to a separation distance that is 1.49$\arcsec$ larger than the FWHM of the synthesized beam in that direction.  This provided 62 independent calculations of Equations (4) and (5).  To check for bias from this choice of spacing, larger spacings of 10 pixels were tried and consistent results were obtained.  Differentiation was performed using the Savitzky-Golay method \cite{sg64}.  Integration was performed by summing over a row of pixels in a map of the integrand.  

\vskip 15ex

\section{STATISTICS}%3

\subsection{Uncertainties}%3.1

The analysis of the results rely on estimates of the uncertainties for the coefficients and the total uncertainties for calculations of Equations (4) and (5).  The latter are not shown explicitly in this paper, but are part of the chi-squared ($\chi^{2}$; and reduced chi-squared, $\chi^{2}_{\nu}$) values used in the analysis.  How these uncertainties were estimated is explained below.  
%f3
\begin{figure}[t!]
\centering
\includegraphics[width=0.47\textwidth]{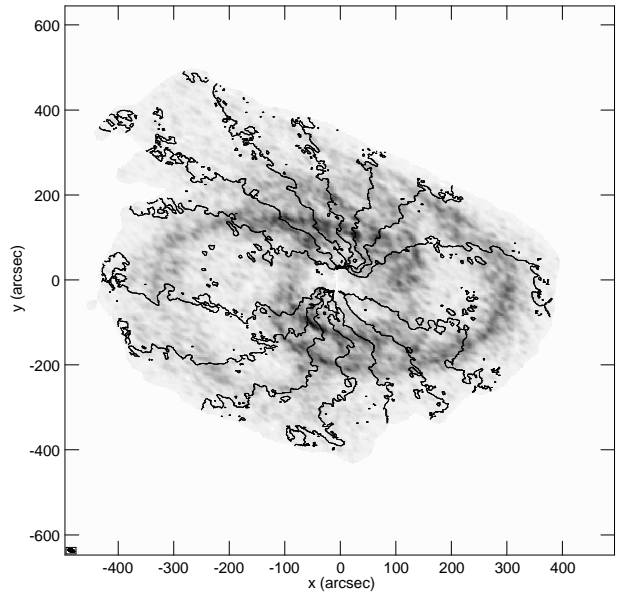} 
\caption{Rotated H{\hskip 1.5pt \footnotesize  I} specific intensity map of NGC 1365 overlaid with contours of constant $v_y$.   The galaxy is oriented in the same way as shown in Figure 1.  The central, most vertical contour is $v_y$ = 0 km s$^{-1}$.  Velocity contours are spaced 80 km s$^{-1}$ apart.  The receding half of the galaxy is to the right.  The peak flux is 250 Jy m s$^{-1}$ beam$^{-1}$.  In the bottom left corner is a plot of the synthesized beam.  Note that the H{\hskip 1.5pt \footnotesize  I} does not trace the bar pattern.} 
\end{figure}

The uncertainties for the coefficients are reported as the half-width, HW, of their 95$\%$ confidence intervals.  The 95$\%$ confidence interval for the $i$th coefficient is,
%e16
\begin{equation}
 t_\nu(2.5\%)\mbox{SE}_i \leqslant \alpha_i \leqslant t_\nu(97.5\%)\mbox{SE}_i,
 \end{equation}
where $t_\nu$ is the percentile of a Student's $t$ - distribution with $\nu$ degrees of freedom, and SE$_i$ is the standard error for the $i$th coefficient (see Ramsey \& Shafer 2002).  The solutions in this paper have enough degrees of freedom that $t_\nu$(2.5\%) = $t_\nu$(97.5\%).  The HW of a 95$\%$ confidence interval for the $i$th coefficient is,
%e17
\begin{equation}
\mbox{HW}_i =  t_\nu(97.5\%)\mbox{SE} _i.
\end{equation}
In a similar fashion, the HW of a 95\% confidence band for $\Omega_p$ when Equations (2) or (3) are summed to order $n$ $\geqslant$ 1 is,
%e18
\begin{equation}
\mbox{HW}(r) = t_\nu(97.5\%)\mbox{SE}(r),
\end{equation}
where SE($r$) is the standard error of $\Omega_p$ as a function of radius.  For the solutions in this paper $t_\nu$(97.5\%) $\approx$ 2.

The standard errors were found using Monte Carlo methods.  Monte Carlo methods are preferred to those using the covariance matrix because such methods assume $\sigma_{G_{j,i}}$ = 0, which is not true in this case.  For ten-thousand iterations, normally distributed random noise with standard deviations of $\sigma_{G_{j,i}}$ and $\sigma_{d_j}$ were added to {\bf G} and {\bf d} respectively, and solutions were found for the coefficients.  The standard deviations in the distributions of the results were adopted as estimates of SE$_i$.  Likewise, the standard deviations for distributions of $\Omega_p$ over a range of $r$ were adopted as estimates of SE($r$).  Monte Carlo methods produced confidence intervals and confidence bands $\sim$ 1 - 10\% greater than methods using the covariance matrix.  The difference was greatest for forms of $\Omega_p$ having the most terms.

Estimates of $\sigma_{G_{j,i}}$ and $\sigma_{d_j}$ were found by propagating estimates of the uncertainty per pixel, $\sigma_I$ and $\sigma_{v_y}$ given in Section 2.3, as well as the uncertainties in $x$ and $y$, through the calculations of {\bf G} and {\bf d}.  The RMS noise in a line-free channel is a lower limit to $\sigma_{I}$ in a map of $I$ where emission is present, so doubling the RMS noise should adequately account for $\sigma_{I}$.  To estimate $\sigma_{v_y}$, the uncertainty per pixel in velocity resolution, the uncertainty in $V_{sys}$, and the uncertainty in inclination were propagated through the calculation of $v_y$.  Following the reasoning of \cite{r74}, the uncertainty per pixel that is due to the velocity resolution was estimated as the channel width divided by the number of channels in which a detection occurs.  JvM95 show that only the brightest H{\hskip 1.5pt \footnotesize  I} appears in more than two channels of the data cube, the weakest detections appear in one, and the majority appears in about two.  The uncertainty due to the velocity resolution was therefore set to one-half the channel width.  The uncertainties in $x$ and $y$ were calculated from the uncertainties in the kinematic center and the position angle (PA).  The uncertainty in $y$ also included the uncertainty in the inclination angle.

To avoid the need for propagating the uncertainties in $x$ and $y$ that are due to uncertainties in the distance to NGC 1365, the instrumental units of the VLA, [$\Omega_p$] = km s$^{-1}$ arcsec$^{-1}$, are used for calculating Equations (4) and (5) and reporting the results.  Furthermore, using units common for galactic disks, [$\Omega_p$] = km s$^{-1}$ kpc$^{-1}$, would introduce an unnecessary amount of uncertainty when determining the functional form of $\Omega_p$ that provides the best solution, or when comparing the results with those obtained using other solution methods and from previous estimates.  After the analysis, the most important results are reported in units commonly used for galactic disks.  The uncertainties for the converted results are reported as 1$\sigma$ uncertainties, which were estimated by propagating the standard errors for the coefficients and the uncertainty for the distance through the conversion.   

Errors in the PA can contribute to errors in the coefficients that are not represented in the propagated uncertainties.  Monte-Carlo error estimations were used to examine this effect.  For ten-thousand iterations, a PA was randomly chosen from a normal distribution of angles centered on the assumed PA whose standard deviation was equal to the uncertainty in PA, $\sigma_{PA}$.  The resulting distributions in the coefficients were typically non-gaussian, and produced pattern speeds that were most often smaller than those derived using the assumed PA.  It would therefore be inappropriate to add the standard deviations in the results for the Monte-Carlo estimates in quadrature to the uncertainties in the coefficients in order to account for $\sigma_{PA}$.  Instead, the effect of an incorrect PA is discussed for each galaxy on a case by case basis.  The percent error due to an incorrect PA is much smaller than those that are typically reported for the original solution method of TW84.  The percent error for the original solution method can be as large as 25\% for a 2$^{\circ}$ error in PA, and 100\% for a 5$^{\circ}$ error in PA (Debattista 2003; Debattista \& Williams 2003).  The precent errors for the current method are about 1/4 as large.   

Finally, the total uncertainty for each $j$th calculation of Equation (4) or (5), $\sigma_j$,  was used for finding $\chi^2$ (and $\chi^{2}_{\nu}$).  These were estimated as, 
%e19
\begin{equation}
\sigma^{2}_j = \sigma^{2}_{d_j} + \sum_{i=0}^{n}\sigma_{G_{j,i}}^{2}\alpha_i^{2},
\end{equation}	
by following the reasoning of Bevington \& Robinson (2002) to take into account both $\sigma_{d_j}$ and $\sigma_{G_{j,i}}$.  

\subsection{Analysis}%3.2

The primary goal of the analysis is to determine the functional form of $\Omega_p$ that provides the best solution, and thus the best estimate of $\Omega_p$.  Starting at order $n$ = 0, the value of $\chi^{2}_\nu$ was compared with the value obtained after the addition of an extra term.  If there was no improvement in $\chi^{2}_\nu$, then the additional term did not provide an improvement to the solution.  If there was an improvement in $\chi^{2}_\nu$, then the $F$ test of an additional term was used to find the statistical significance of the improvement.  It consists of calculating
%e20
\begin{equation}
F = {{\triangle\chi^{2}}\over{\chi^{2}_\nu}},
\end{equation}
and finding the probability, $P$, of $F$ being as large with the addition of an extra term, should the coefficient of the extra term be zero (Bevington \& Robinson 2002).  A value of $P$ $\lesssim$ $1\%$ is convincing evidence that the extra term is non-zero, a value of $P$ $\sim$ $5\%$ is suggestive, but inconclusive, and larger values are an indication that the improvement is statistically insignificant.  A lower limit of 0.01\% was chosen as a cutoff for reporting $P$.  Such a small percentage means that the improvement in $\chi_\nu^{2}$ is extremely statistically significant.  More information about how to interpret values of $P$ in the $F$ test can be found in \cite{rs02}.  For the $n$ = 1 term, an $F$ test providing convincing evidence that it is non-zero is equivalent to relaxing the assumption that $\Omega_p$ is constant.

Terms were added based on the results of the $F$ test until there was no longer an improvement in $\chi^{2}_\nu$, the improvement was no longer statistically significant, or {\bf G}$^T${\bf G} became ill conditioned for numerical inversion.  The matrix {\bf G}$^T${\bf G} will eventually become ill conditioned for numerical inversion in the same manner that is expected for least squares solutions using polynomials with successively higher order terms.  When this occurs the solution is unstable and the overall statistics are unreliable \cite{a05}.    
 %t2
\begin{deluxetable*}{lccccc}
\tablecaption{Best-Fit Coefficients for NGC 1365} 
\tablewidth{0pt}
\startdata
\tableline\tableline\\[-7.5 pt]
& &$\alpha_0$ $\pm$ HW$_0$&$\alpha_1$ $\pm$ HW$_1$&$\alpha_2$ $\pm$ HW$_2$&$\alpha_3$ $\pm$ HW$_3$ \\
Equation &$n$&(km s$^{-1}$ arcsec$^{-1}$) &10$^{-2}$ (km s$^{-1}$ arcsec$^{-2}$)&10$^{-4}$ (km s$^{-1}$ arcsec$^{-3}$)&10$^{-7}$ (km s$^{-1}$ arcsec$^{-4}$)\\
 \tableline\\[-7.5 pt]
(2), (3) & 0 &  1.27 $\pm$ 0.10 & \nodata & \nodata & \nodata \\[2 pt]
\tableline\\[-7.5 pt]
(2) &1 &  2.68 $\pm$ 0.38 &$-$0.55 $\pm$ 0.13 & \nodata & \nodata \\
&2 &  4.05	 $\pm$ 0.83 &$-$1.65 $\pm$ 0.57  & 0.20 $\pm$ 0.09 & \nodata \\  
&3 &  5.93 $\pm$ 1.43 & $-$4.06 $\pm$ 1.64 & 1.11 $\pm$ 0.59 & $-$1.05 $\pm$ 0.64 \\[2 pt]
\tableline\\[-7.5 pt]
& & 10$^{-1}$ (km s$^{-1}$ arcsec$^{-1}$)  & 10$^{2}$ (km s$^{-1}$) & 10$^{3}$ (km s$^{-1}$ arcsec) &  \nodata \\[2 pt]
\cline{3-6}\\[-7.5 pt]
(3) &1 & $-$0.59 $\pm$ 2.88 & 2.87 $\pm$ 0.61& \nodata & \nodata \\
& 2 & $-$1.04 $\pm$ 4.28 & 3.03 $\pm$ 1.29 & 0.95 $\pm$ 6.65 & \nodata 
\enddata
\end{deluxetable*} 

In addition to the above statistics, $R^2$ was found for a solution when Equation (2) or (3) was summed to order $n$ $\geqslant$ 1.  In this paper $R^2$ is the percent of the variance for the solution when $n$ = 0 that is explained by the solution using more than 1 term.  The value of $R^2$ in combination with the results from the $F$ test were used for comparing the goodness of fit for the two different forms of $\Omega_p$. 

\section{RESULTS AND ANALYSIS}%4

In this section the results and analysis are presented by the equation used for $\Omega_p$, and then using plots.  At the end of this section conclusions are drawn from the analysis.  The results for the best-fit coefficients, $\alpha_i$, and the half-widths of their 95\% confidence intervals, HW$_i$, are shown in Table 2.  The statistics for analyzing the results are shown in Table 3.

\subsection{Results From Equation (2)}%4.1

Solutions using Equation (2) show improvements in $\chi_\nu^{2}$ that are statistically significant to order $n$ = 3 according to the $F$ test ($P$ $<$ 0.01\%).  The overall statistics, however, are only reliable to order $n$ = 2, after which point {\bf G}$^T${\bf G} became ill conditioned for numerical inversion.  The improvement in $\chi_\nu^{2}$ for order $n$ = 1 is extremely statistically significant ($P$ $<$ 0.01\%), and explains more than half of the variance ($R{^2}$ = 59.5\%) for the solution when $n$ = 0.  The improvement in $\chi_\nu^{2}$ for order $n$ = 2 is as statistically significant ($P$ $<$ 0.01\%), and explains more of the variance ($R{^2}$ = 70.1\%) for the solution when $n$ = 0.  Since there is convincing evidence for a radial dependence in $\Omega_p$, the solution when $n$ = 0 is the mean pattern speed.  It follows that the $i$ = 1 coefficient for the $n$ = 1 solution is the mean shear rate, and the $i$ = 2 coefficient for the $n$ = 2 solution is a correction for the radial variance in the shear rate.  

\subsection{Results From Equation (3)}%4.2

The solution using Equation (3) to order $n$ = 1 shows an improvement in $\chi_\nu^{2}$ that is extremely statistically significant according to the $F$ test ($P$ $<$ 0.01\%).  Its value of $F$ = 525 is quite large, and is more than twice the value of $F$ = 241 for the $n$ = 1 solution using Equation (2).  This solution also explains more of the variance ($R^{2}$ = 79\%) for the solution when $n$ = 0 than do the solutions using Equation (2).  The $i$ = 0 term for this solution is indistinguishable from zero given the size of its uncertainty.  To a very good approximation, the $i$ = 0 term is negligible, and the form of this solution is 1/$r$.  The $n$ = 2 solution did not provide a statistically significant improvement in $\chi_\nu^{2}$ ($P$ = 7.57\%). 

\subsection{Plots of the Results}%4.3

%t3
\begin{deluxetable}{lccccc}
\tablecaption{Statistics for NGC 1365} 
\tablewidth{0pt}
\startdata
\tableline\tableline\\[-7.5 pt]
&&&&$P$&$R^2$ \\
Equation&$n$&$\chi_\nu^{2}$&$F$&($\%$)&($\%$)\\
\tableline\\[-7.5 pt]
(2), (3) & 0 &  21.9  &   \nodata & \nodata & \nodata  \\[2 pt]
\tableline\\[-7.5 pt]
(2) &1 & 4.43  &  241 & $<$ 0.01  & 59.5 \\
&2 &  2.30 & 56.7  &   $<$ 0.01 & 70.1 \\  
&3 &  0.72  & 129  &  $<$ 0.01 & 76.3 \\[2 pt]
\tableline\\[-7.5 pt]
(3) &1 &  2.28 & 525  & $<$ 0.01  & 79.0\\
&2 & 2.19  &  3.37  & \hskip 2.6ex7.14  & 79.0
\enddata
\end{deluxetable} 

Plots of $\Omega_p$ from the results in Table 2 are more revealing.  Figure 4 shows plots of $\Omega$ adopted from the rotation curve of JvM95, possible locations for 2 and 4 arm Lindblad resonances, and $\Omega_p$ from the results for the solution using Equation (2) to order $n$ = 2.  The purpose of this figure is to build a complex plot one layer at a time, the format of which is used in subsequent plots of the results.  The uncertainties for $\Omega$ are omitted because they are as small or smaller than the circles representing it.  Results are not plotted for $r$ $<$ 50$\arcsec$, wherein there is not an H{\hskip 1.5pt \footnotesize  I} pattern. 
 
Figure 5 shows plots for all of the results in Table 2.  Panel (a) shows the result when $n$ = 0.  Panels (b), (c), and (d) show pattern speeds having the form of Equation (2) to order $n$ = 1, 2, and 3 respectively.  Panels (e) and (f) show pattern speeds having the form of Equation (3) to order $n$ = 1 and  2 respectively.  Upon comparison with the other solutions, the solution in panel (a) is clearly  the mean pattern speed.  The larger confidence bands in panel (d) for the $n$ = 3 solution using Equation (2) are a result of the instability of that solution.  There is hardly any noticeable difference between panels (e) and (f).  Note the progression in panels (a) thorough (d) where the form of $\Omega_p$ converges to a form resembling that in panels (e) and (f).  As terms are added to Equation  (2), the radial behavior of $\Omega_p$ becomes increasingly more concave up.  The same behavior is achieved by Equation (3) with fewer terms.   

Plots of the results do not show the pattern speed of the bar.  There are enough terms in the forms of $\Omega_p$ in panels (d) and (f) to produce a turnover that is concave down around 100$\arcsec$-150$\arcsec$, inside of which would be a constant $\Omega_p$ for the bar.  Although NGC 1365 is often regarded as a typical case for a continuous bar and spiral pattern having the same pattern speed, the bar pattern speed may be decoupled from the spiral arm pattern speed (Sellwood \& Sparke 1988, Sellwood 1993 and references therein).  

\begin{figure*}
\centering
\includegraphics[width=0.94\textwidth]{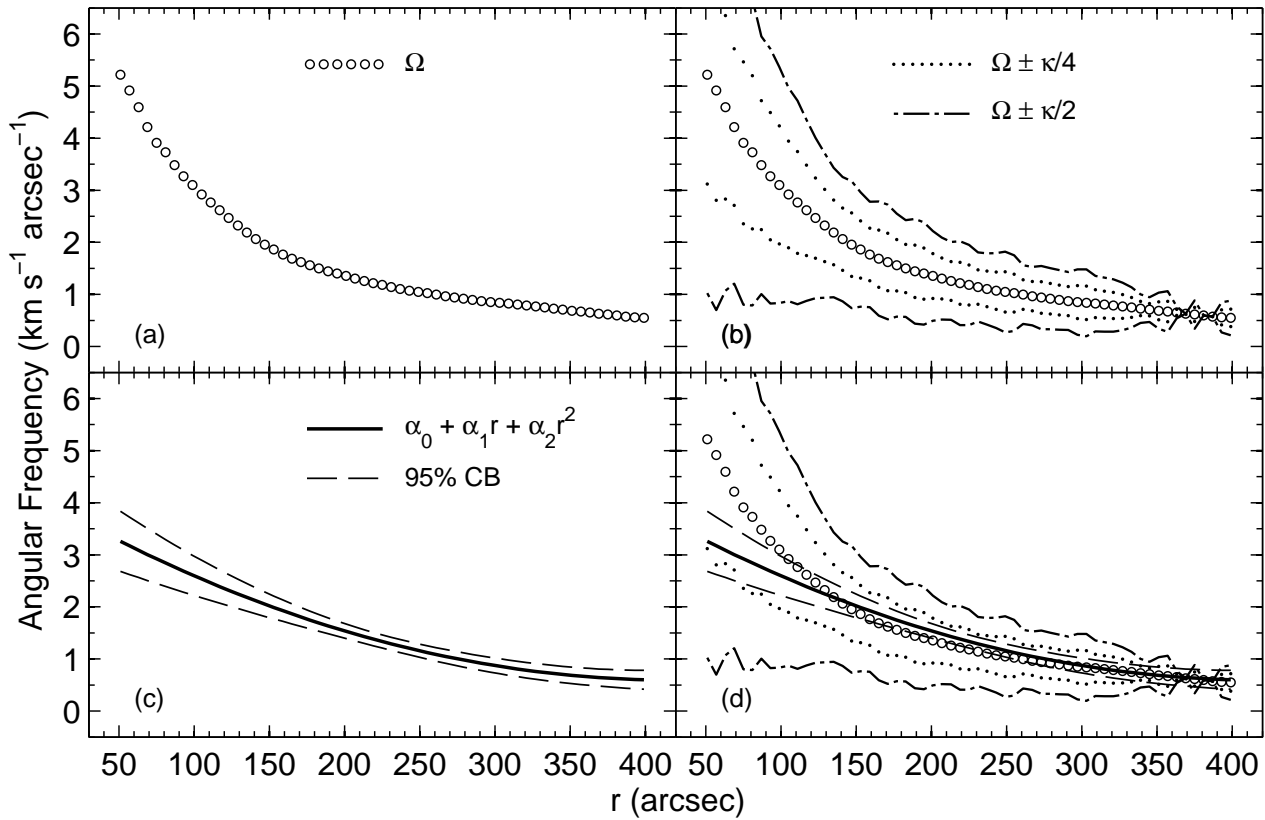}
\caption{Plots for explaining how subsequent plots of the results are formatted.  Panel (a) shows the material speed, $\Omega$, represented by circles.  Panel (b) shows $\Omega$ with possible locations for Lindblad resonance.  Dotted and dash-dot lines represent possible locations for 2 and 4 arm Lindblad resonances respectively.  Panel (c) shows the pattern speed, $\Omega_p$, using the results for Equation (2) to order $n$ = 2 and its 95\% confidence bands.  These are represented by solid and dashed lines respectively.  Panel (d) is a combination of panels (b) and (c).  See text for more details.} 
\end{figure*}
%f5
\begin{figure*}
\centering
\includegraphics[width=0.94\textwidth]{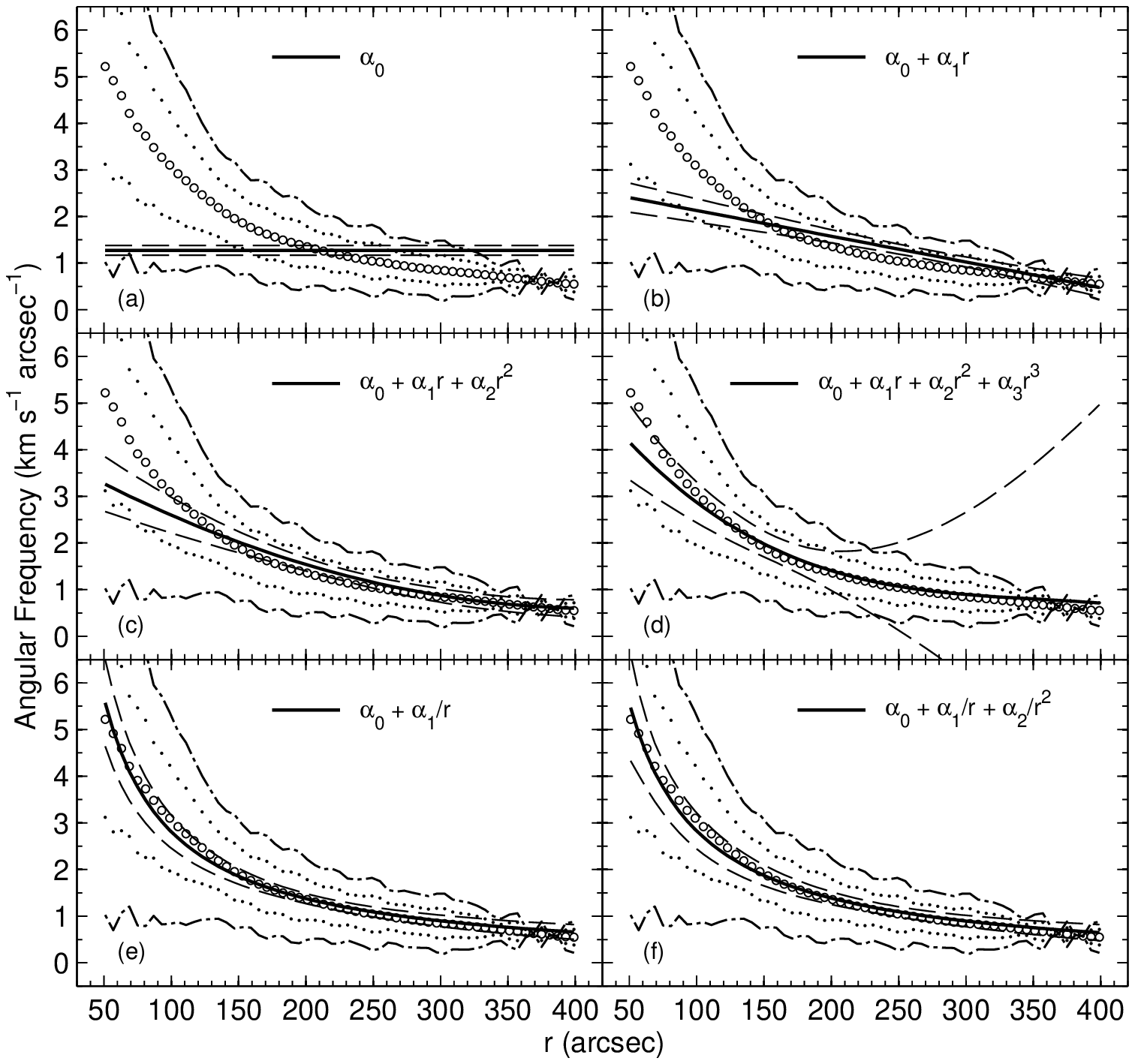} 
\caption{Plots of pattern speeds from the results in Table 2.  They are formatted in the same way as panel (d) in Figure 4.  See text for details.  Note the similarity of panels (d), (e), and (f), with larger confidence bands in panel (d).  The material speed fits within the 95\% confidence bands in panels (e) and (f).} 
\end{figure*} 

To explore the effect of an incorrect PA, solutions were found for a range of PA within $\pm$ 2$\sigma_{PA}$ of the assumed PA.  The results for the mean pattern speed and the values of $F$ for the $n$ = 1 solutions are shown in Figure 6.  The changes in the mean pattern speed are indicative of the changes for all forms of $\Omega_p$ in Figure 5.  When the mean pattern speed decreases or increases by a certain amount, the other forms generally do so as well.  If the assumed PA is incorrect, the results are most likely over estimating $\Omega_p$.  The difference could be as large as $\sim$ 0.2 km s$^{-1}$ arcsec$^{-1}$ (or $\sim$ 2.3 km s$^{-1}$ kpc$^{-1}$ using the assumed distance stated in Section 4.4).  The mean percent difference within $\pm$ 1$\sigma_{PA}$ is 7\%.  The largest percent difference within $\pm$ 1$\sigma_{PA}$ is $\sim$ 18\%.  For all of the changes in PA shown, there is convincing evidence for shear in the pattern, and the values of $F$ are the largest for $\Omega_p$ having form of Equation (3).  

 %%f6
\begin{figure}
\centering
\includegraphics[width=0.47\textwidth]{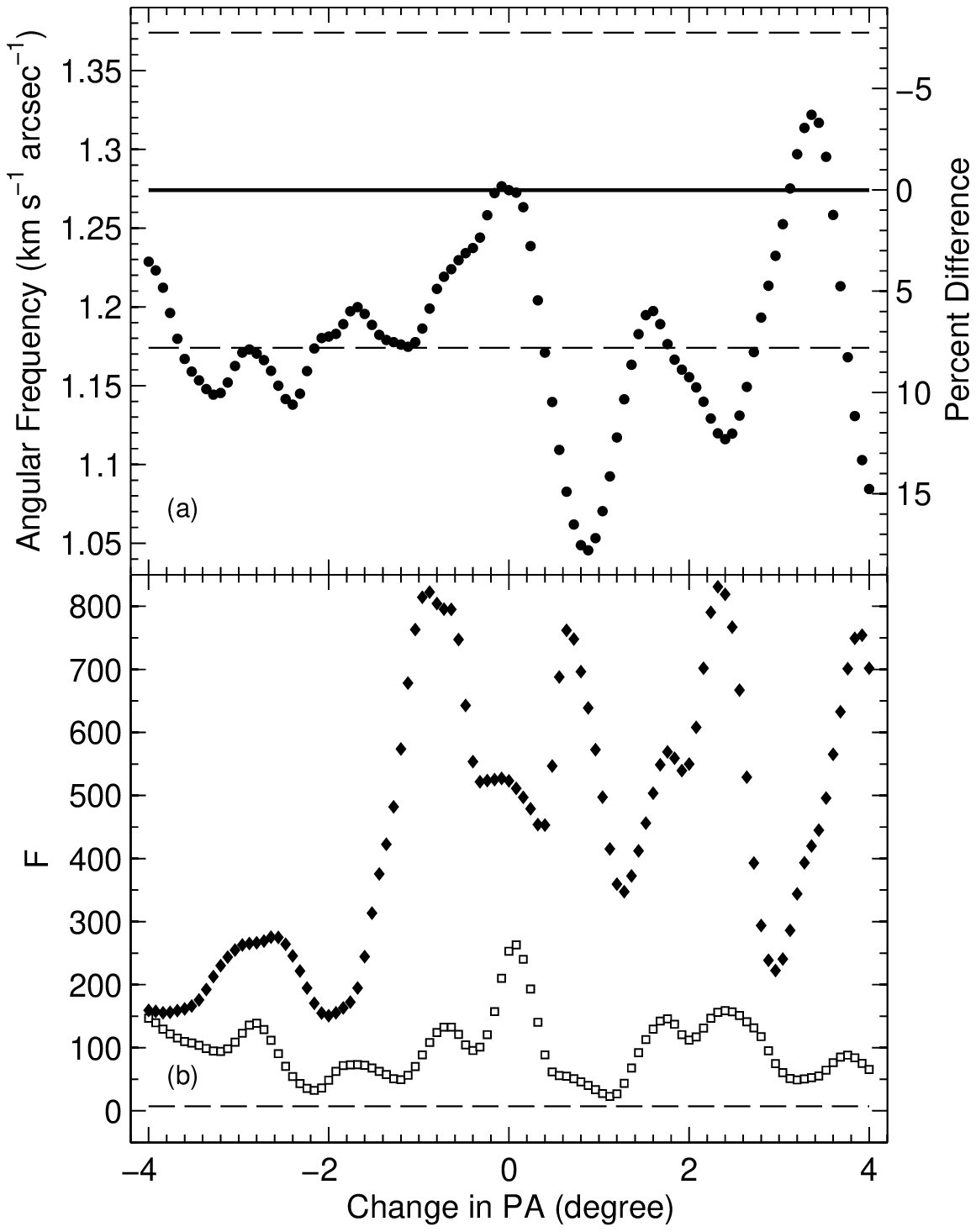}
\caption{Effect of an incorrect PA for NGC 1365.  Panel (a) shows how the mean pattern speed, represented by points, varies with PA.  For reference, the solid and dashed lines in panel (a) represent the mean pattern speed and its 95\% confidence interval obtained using the assumed PA.  Panel (b) shows the values of $F$ for the $n$ = 1 solution as open squares for Equation (2) and solid diamonds for Equation (3).  The dashed line in panel (b) is the value of $F$ at which $P$ = 1\%.} 
\end{figure}

\subsection{Conclusions}%4.4

The best estimate of $\Omega_p$ is adopted from the results for Equation (3) to order $n$ = 1.  This conclusion is based on three features of the results: 1) this solution has the greatest values of $F$ and $R^2$; 2) the results for Equation (2) converge towards a form resembling this solution as successive terms are added; and 3) this solution has a simpler form than the $n$ = 2 solution using Equation (2).  Although inferring the exact functional form of $\Omega_p$ is beyond the scope of the results, its radial behavior is approximately 1/$r$.  

There are no clear indications of resonances associated with the spiral arms.   Panel (e) in Figure 5 does not show a single point of corotation resonance.  In the same panel, Lindblad resonances can neither be confirmed nor ruled out beyond 350$\arcsec$.  The intersection of $\Omega_p$ and the possible locations for Lindblad resonances beyond 350$\arcsec$ may be a coincidence due to the precision of the results.  The results exclude nonlinear mode coupling of the bar and spiral arms, which require inner Lindblad resonances where the spiral arms begin (Tagger at al. 1997; Masset \& Tagger 1997).  The effect of an incorrect PA would not increase $\Omega_p$ enough to account for the lack of inner Lindblad resonances at the beginnings of the spiral arms.

%t4
\begin{deluxetable}{lcr@{.}l}
\tablecaption{Results for NGC 1365 in Common Units} 
\tablewidth{0pt}
\startdata
\tableline\tableline\\[-7.5 pt]
Mean pattern speed&& 14&5 ($\pm$ 2.16) km s$^{-1}$ kpc$^{-1}$ \\
Mean shear rate&& $-$ 0&71 ($\pm$ 0.24) km s$^{-1}$ kpc$^{-2}$ \\
Best estimate of $\Omega_p$&& 2&87 ($\pm$ 0.61) $\times$ 10$^{2}$ km s$^{-1}$ /$r$ 
\enddata
\end{deluxetable}  
   
The most important results are summarized in Table 4 using common units.  The units were converted using an assumed distance of 18.1 ($\pm$ 2.6) Mpc to NGC 1365, which was estimated from the mean of 37 redshift-independent measurements found in the Nasa Extragalactic Database (NED).  The 1$\sigma$ uncertainties are shown in parenthesis.  About half of the 1$\sigma$ uncertainties are due to the uncertainty in the distance. 

\section{A TEST FOR SELECTION BIAS}%5

There is a possibility that the results for NGC 1365 are biased by selecting the region $|y|$ $\leqslant$ 400$\arcsec$ (hereafter region $\mathcal{A}$) for applying the method.  To test for whether this occurred, additional solutions were found for the regions $|y|$ $\leqslant$ 200$\arcsec$, $-$400$\arcsec$ $\leqslant$ $y$ $\leqslant$ 0$\arcsec$, and 0$\arcsec$ $\leqslant$ $y$ $\leqslant$ 400$\arcsec$ (hereafter regions $\mathcal{B}$, $\mathcal{C}$, and $\mathcal{D}$ respectively).  Solutions for region $\mathcal{B}$ use less of the H{\hskip 1.5pt \footnotesize  I} in the outer part of the disk than solutions for region $\mathcal{A}$.  Comparing the results for these two regions will show whether $\Omega_p$ behaves much differently in the inner and outer parts of the disk.  Comparing the results for regions $\mathcal{C}$ and $\mathcal{D}$ will show whether $\Omega_p$ behaves much differently for either side of the galaxy.  The results for regions $\mathcal{B}$, $\mathcal{C}$, and $\mathcal{D}$ are shown in Table 5 up to the same order as the results for region $\mathcal{A}$ in Table 2.  

%t5
\begin{deluxetable*}{lllcccc}
\tablecaption{Best-Fit Coefficients for Different Regions of NGC 1365} 
\tablewidth{0pt}
\startdata
\tableline\tableline\\[-7.5 pt]
&&&$\alpha_0$ $\pm$ HW$_0$&$\alpha_1$ $\pm$ HW$_1$&$\alpha_2$ $\pm$ HW$_2$&$\alpha_3$ $\pm$ HW$_3$\\
Equation&$n$&Region&(km s$^{-1}$ arcsec$^{-1}$)&10$^{-2}$ (km s$^{-1}$ arcsec$^{-2}$)&10$^{-4}$ (km s$^{-1}$ arcsec$^{-3}$)&10$^{-7}$ (km s$^{-1}$ arcsec$^{-4}$)\\
\tableline\\[-7.5 pt]
(2), (3)&0&$\mathcal{B}$& 1.37 $\pm$ 0.12 & \nodata & \nodata & \nodata \\
&&$\mathcal{C}$&1.20 $\pm$ 0.18 & \nodata & \nodata & \nodata \\
&&$\mathcal{D}$&1.31 $\pm$ 0.13 & \nodata & \nodata & \nodata \\[2 pt]
\tableline\\[-7.5 pt]
(2)&1&$\mathcal{B}$& 2.75 $\pm$ 0.50 & $-$0.57 $\pm$ 0.20 & \nodata & \nodata \\
&&$\mathcal{C}$&3.23 $\pm$ 0.66 & $-$0.74 $\pm$ 0.23 & \nodata & \nodata \\
&&$\mathcal{D}$& 2.39 $\pm$ 0.45 & $-$0.44 $\pm$ 0.16 & \nodata & \nodata \\[2 pt]
\cline{2-7}\\[-7.5 pt]
&2&$\mathcal{B}$& 4.65 $\pm$ 1.20 &$-$2.27 $\pm$ 1.02 & 0.34 $\pm$ 0.20 & \nodata \\  
&&$\mathcal{C}$& 4.42 $\pm$ 1.38 &$-$1.86 $\pm$ 1.17 & 0.23 $\pm$ 0.24 & \nodata \\  
&&$\mathcal{D}$&3.81 $\pm$ 1.44&$-$1.50 $\pm$ 0.97& 0.17 $\pm$ 0.14& \nodata \\[2 pt]
\cline{2-7}\\[-7.5 pt]
&3&$\mathcal{B}$& 7.12 $\pm$ 2.10 & $-$5.95 $\pm$ 2.86 & 1.99 $\pm$ 1.24 & $-$2.28 $\pm$ 1.74 \\
&&$\mathcal{C}$ &5.37 $\pm$ 2.18 & $-$3.25 $\pm$ 2.93 & 0.83 $\pm$ 1.25& $-$0.81 $\pm$ 1.70\\
&&$\mathcal{D}$ &6.35 $\pm$ 2.28 & $-$4.62 $\pm$ 2.39 & 1.31 $\pm$ 0.81 & $-$1.27 $\pm$ 0.87\\[2 pt]
\tableline\\[-7.5 pt]
&&&10$^{-1}$ (km s$^{-1}$ arcsec$^{-1}$)&10$^{2}$ (km s$^{-1}$)&10$^{3}$ (km s$^{-1}$ arcsec)&  \nodata \\[2 pt]
\cline{4-7}\\[-7.5 pt]
(3)&1&$\mathcal{B}$&\hskip 2ex0.40 $\pm$ 3.66 & 2.70 $\pm$ 0.71 & \nodata & \nodata \\
&&$\mathcal{C}$& $-$0.06 $\pm$ 3.83 & 2.78 $\pm$ 0.78 & \nodata & \nodata \\
&&$\mathcal{D}$&$-$1.39 $\pm$ 4.53& 3.03 $\pm$ 0.97 & \nodata & \nodata \\[2 pt]
\cline{2-7}\\[-7.5 pt]
&2&$\mathcal{B}$& \hskip 2ex0.55 $\pm$ 5.34 & 2.67 $\pm$ 1.50 & 0.17 $\pm$ 7.18 & \nodata \\
&&$\mathcal{C}$& $-$1.25 $\pm$ 6.17 & 3.23 $\pm$ 1.98 & 2.35 $\pm$ 9.28 & \nodata \\
&&$\mathcal{D}$&$-$0.74 $\pm$ 6.41 & 2.79 $\pm$ 2.03 & 1.75 $\pm$ 13.4 & \nodata 
\enddata
\end{deluxetable*} 

The coefficients for regions $\mathcal{B}$, $\mathcal{C}$, and $\mathcal{D}$ are consistent with those for region $\mathcal{A}$.  This is easily seen by comparing plots of $\Omega_p$, which are shown in Figures 7, 8, and 9 for regions $\mathcal{B}$, $\mathcal{C}$, and $\mathcal{D}$ respectively.  They show the same progression as was observed in Figure 5 where that the form of $\Omega_p$ in panels (a) through (d) converges to a form resembling that in panels (e) and (f).  Any differences among Figures 5, 7, 8, and 9 are very small.  Compared with Figure 5, the plotted pattern speeds in Figures 7, 8, and 9 have slightly larger confidence bands.  This is due to having fewer calculations of Equations (4) and (5).   

%f7
\begin{figure*}
\centering
\includegraphics[width=0.94\textwidth]{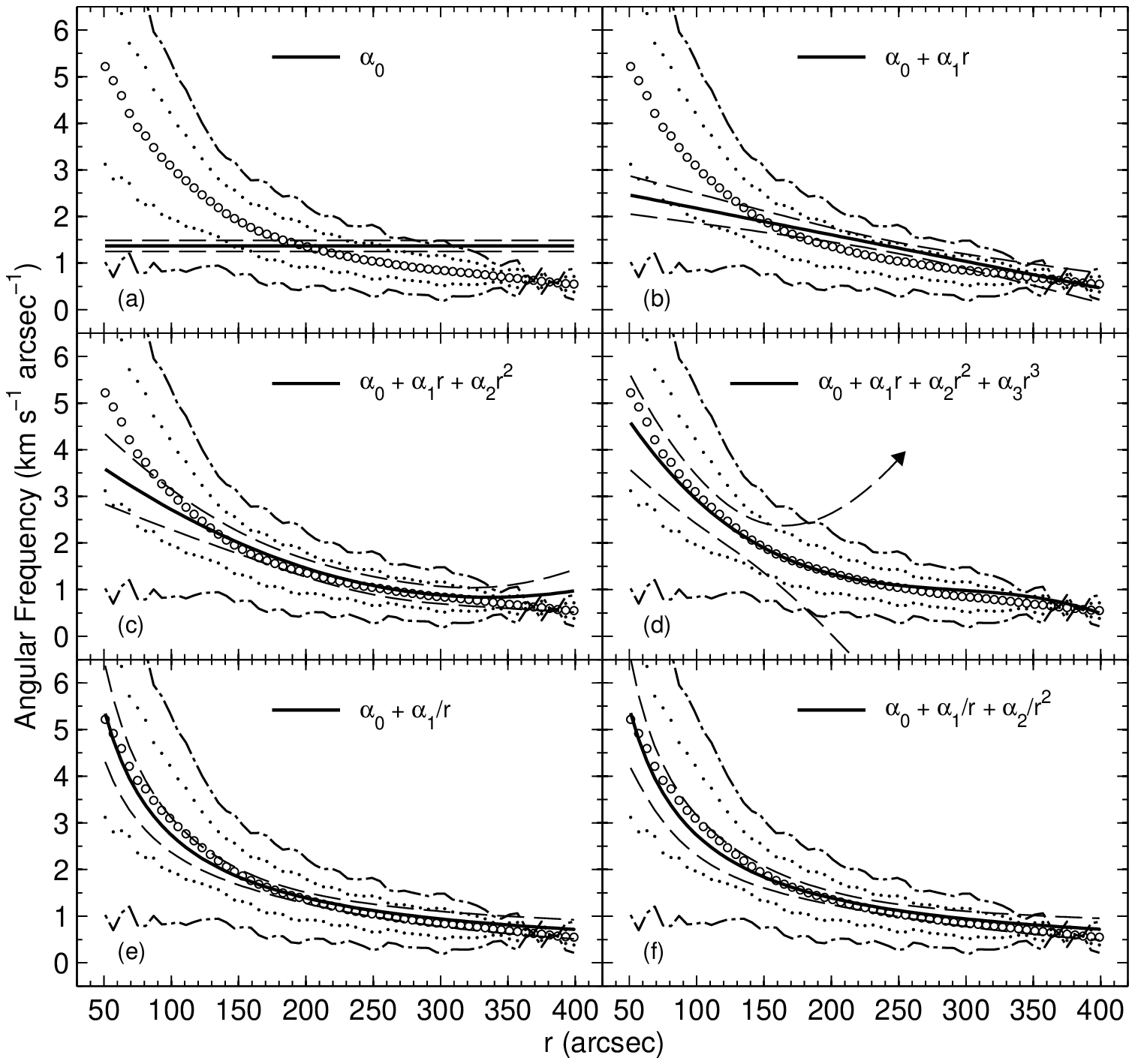} 
\caption{Plots of pattern speeds for region $\mathcal{B}$.  The figure is formatted the same way as Figure 5.  In panel (d) the arrow at the end of the upper confidence band points in the direction it continues.} 
\end{figure*}
%f8
\begin{figure*}
\centering
\includegraphics[width=0.94\textwidth]{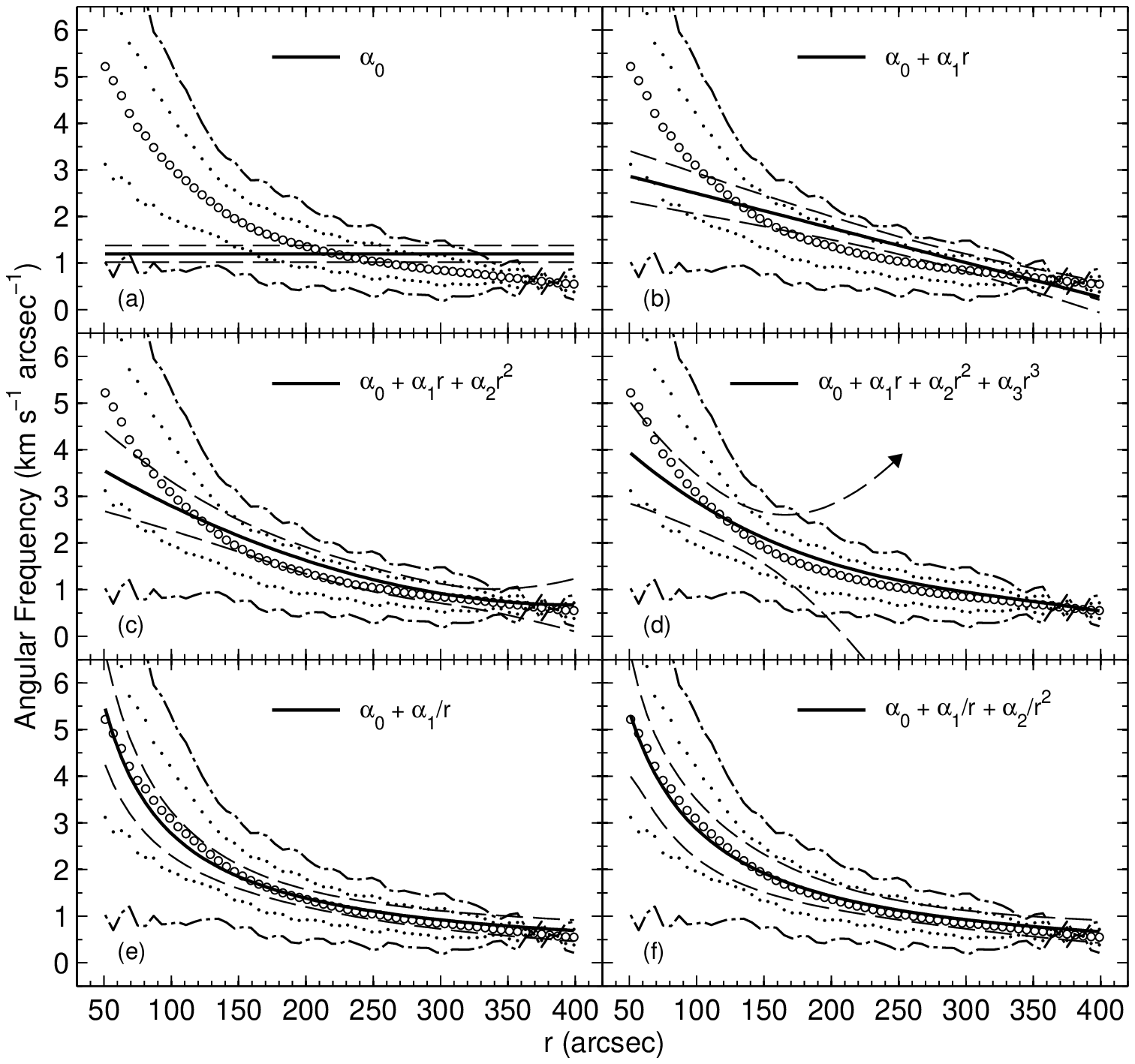} 
\caption{Plots of pattern speeds for region $\mathcal{C}$.  The figure is formatted the same way as Figure 5.  In panel (d) the arrow at the end of the upper confidence band points in the direction it continues.} 
\end{figure*}
%f9
\begin{figure*}
\centering
\includegraphics[width=0.94\textwidth]{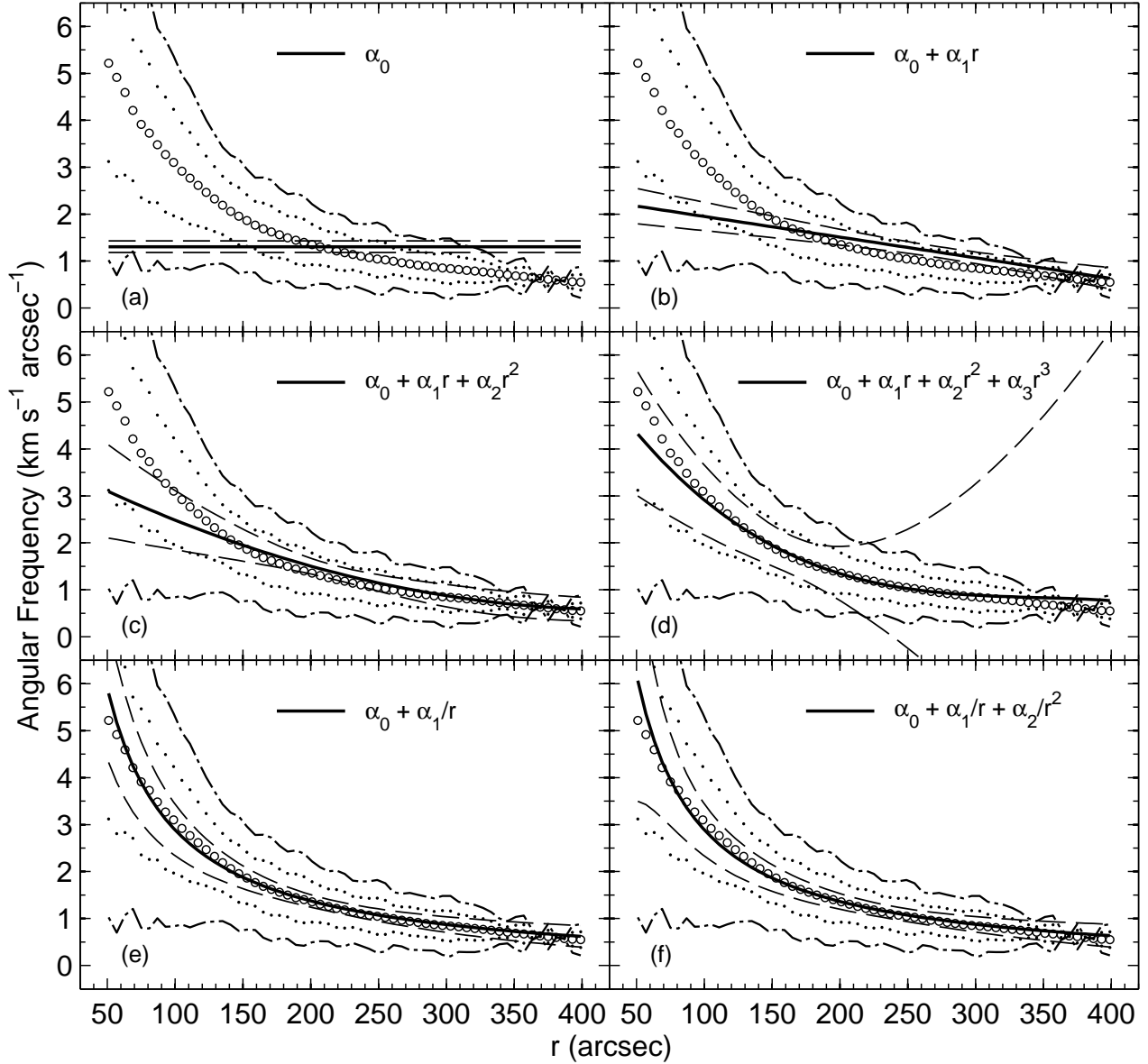} 
\caption{Plots of pattern speeds for region $\mathcal{D}$.  The figure is formatted the same way as Figure 5.} 
\end{figure*}
%f10
\begin{figure*}
\centering
\includegraphics[width=0.80\textwidth]{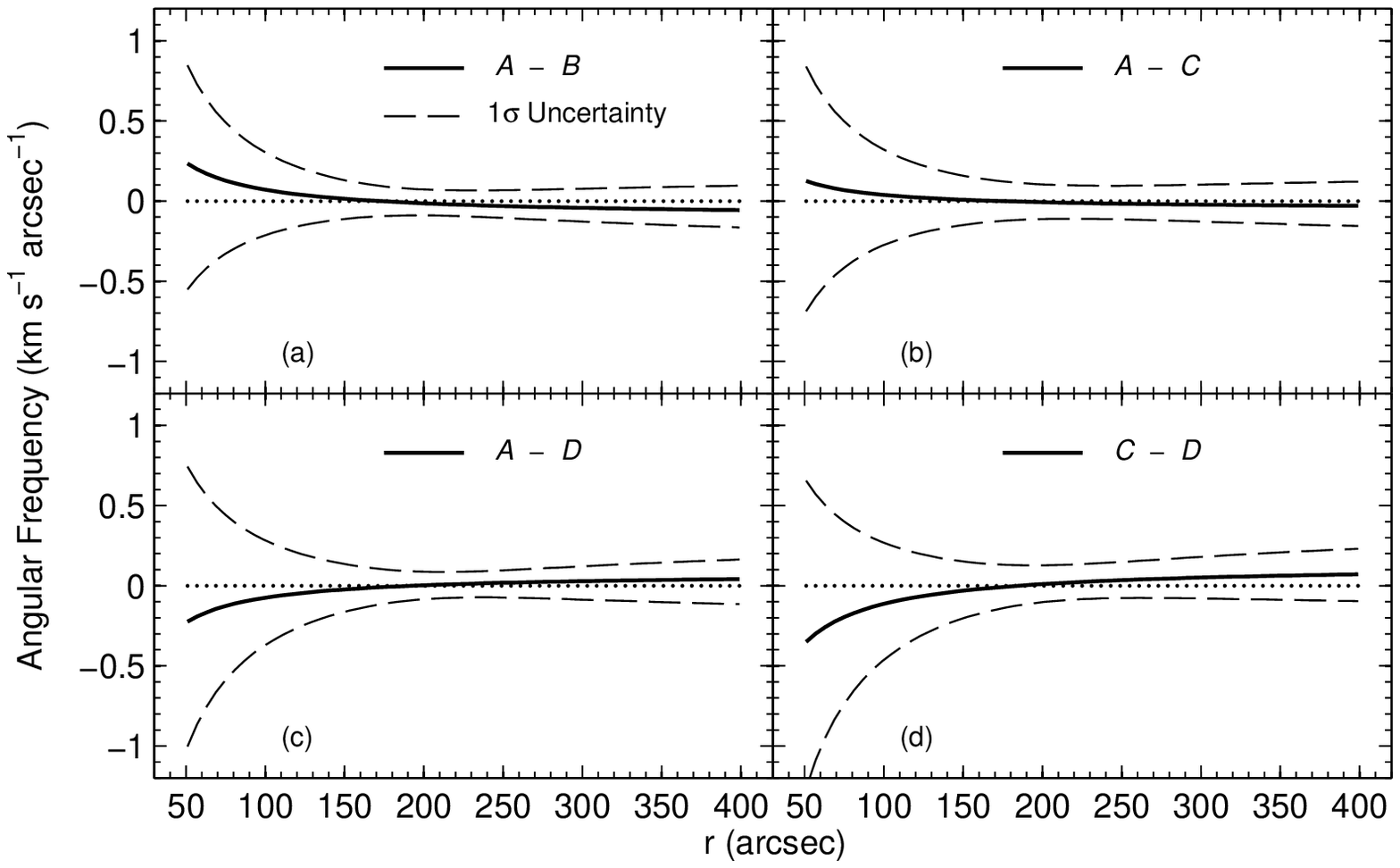} 
\caption{Plots for comparing $\Omega_p$ among regions $\mathcal{A}$, $\mathcal{B}$, $\mathcal{C}$, and $\mathcal{D}$, where that $\Omega_p$ has the form of Equation (3) to order $n$ = 1.  The solid line is the difference between two regions.  The dashed line is the 1$\sigma$ uncertainty for the difference.  The dotted line at zero is provided for reference.  See text for details.} 
\end{figure*}

To see whether the small differences in the plots are significant, the pattern speeds in panel (e) of the figures were subtracted from each other and plotted.  Figure 10 shows four cases to consider: in panel (a) is the result for region $\mathcal{B}$ subtracted from the result for region $\mathcal{A}$; in panel (b) is the result for region $\mathcal{C}$ subtracted from the result for region $\mathcal{A}$; in panel (c) is the result for region $\mathcal{D}$ subtracted from the result for region $\mathcal{A}$; and in panel (d) is the result for region $\mathcal{D}$ subtracted from the result for region $\mathcal{C}$.  The half-widths of the confidence bands shown in the plots are 1$\sigma$ uncertainties that were estimated by adding the standard errors in quadrature.  The largest differences, as well as the largest uncertainties, occur for smaller radii.  The plots show that the differences are less than 10\%, but given that zero falls well within the confidence bands, they are indistinguishable from zero, and therefore insignificant.

%t6
\begin{deluxetable}{llccccc}
\tablecaption{Statistics for Different Regions of NGC 1365} 
\tablewidth{0pt}
\startdata
\tableline\tableline\\[-7.5 pt]
&&&&&$P$&$R^2$ \\
Equation&$n$&Region&$\chi_\nu^{2}$&$F$&($\%$)&($\%$) \\
\tableline\\[-7.5 pt]
(2), (3)& 0 &$\mathcal{B}$ &  5.58 & \nodata & \nodata & \nodata  \\
&&$\mathcal{C}$ &  7.59 & \nodata & \nodata & \nodata  \\
&&$\mathcal{D}$ & 38.6 & \nodata & \nodata & \nodata  \\[2 pt]
\tableline\\[-7.5 pt]
(2)&1& $\mathcal{B}$  & 2.39 & 39.7 & $<$ 0.01 & 52.6 \\
&& $\mathcal{C}$  & 1.86 & 93.6 & $<$ 0.01 & 79.2 \\
&& $\mathcal{D}$ & 3.34 & 317 & $<$ 0.01& 46.6\\[2 pt]
\cline{2-7}\\[-7.5 pt]
&2& $\mathcal{B}$  & 1.38 & 21.4 & $<$ 0.01 & 68.4 \\
&& $\mathcal{C}$  & 1.29 & 13.9 & \hskip 2.6ex0.09 & 85.4 \\
&& $\mathcal{D}$ & 3.21& 2.23 & 14.7 & 55.7 \\[2 pt]
\cline{2-7}\\[-7.5 pt]
&3& $\mathcal{B}$  & 0.62 & 34.0 & $<$ 0.01 & 75.0 \\
&&$\mathcal{C}$  & 0.80 & 18.0 &\hskip 2.6ex 0.02 & 87.0 \\
&& $\mathcal{D}$ & 0.74 & 94.1 & $<$ 0.01 & 68.7 \\[2 pt]
\tableline\\[-7.5 pt]
(3)&1& $\mathcal{B}$  & 1.14 & 114 & $<$ 0.01 & 76.5 \\
&& $\mathcal{C}$ & 1.97 & 86.5& $<$ 0.01 & 86.0 \\
&& $\mathcal{D}$ & 2.22 & 493 & $<$ 0.01& 71.6 \\[2 pt]
\cline{2-7}\\[-7.5 pt]
&2& $\mathcal{B}$  & 1.18 & \nodata & \nodata & 76.5 \\
&& $\mathcal{C}$  &1.98 &\nodata & \nodata & 86.5 \\
&&$\mathcal{D}$ & 2.55 & \nodata& \nodata & 71.7 
\enddata
\end{deluxetable} 

The statistics for the solutions using regions $\mathcal{B}$, $\mathcal{C}$, and $\mathcal{D}$ are shown in Table 6 for comparison with those for region $\mathcal{A}$ shown in Table 3.  Like those in Table 3, the statistics in Table 6 for the solutions using Equation (2) to order $n$ = 3 are unreliable because at this point {\bf G}$^T${\bf G} became ill-conditioned for numerical inversion.  All of the $F$ test results for region $\mathcal{B}$ are consistent with those for region $\mathcal{A}$.  The outer disk is not biasing the analysis of the results.  The $F$ test results for regions $\mathcal{A}$, $\mathcal{C}$, and $\mathcal{D}$ are consistent with each other up to order $n$ = 1 for Equation (2), and up to order $n$ = 2 for Equation (3).  For region $\mathcal{C}$, the $n$ = 2 solution for Equation (2) did not provide a statistically significant improvement in $\chi_\nu^{2}$ ($P$ = 6.78\%).  For region $\mathcal{D}$ the $n$ = 2 solution for Equation (2) did not provide an improvement in $\chi_\nu^{2}$.  Considering that the statistics for region $\mathcal{B}$ are consistent with those for region $\mathcal{A}$, these differences show that the inner disk contributes the most to the concave up behavior of $\Omega_p$ observed in the plotted pattern speeds.  Despite the noted differences, the radial behavior of $\Omega_p$ is approximately the same for both sides of the galaxy.

There are two consistencies among all four regions that are worth emphasizing.  The improvement in $\chi_\nu^{2}$ for solutions using Equations (2) and (3) to order $n$ = 1 is extremely statistically significant for all four regions ($P$ $<$ 0.01\%).  Furthermore, the values of $F$ and $R^{2}$ for solutions using Equation (3) to order $n$ = 1 are the largest among the results that provided a statistically significant improvement in $\chi_\nu^{2}$.  The existence of shear in the pattern, or the approximate 1/$r$ behavior of $\Omega_p$, is not biased by selecting region $\mathcal{A}$ for applying the method.

\vskip 25pt

\section{A TEST TO VERIFY THAT THE METHOD IS NOT MEASURING $\Omega$}%6

The best estimate of $\Omega_p$ is very similar to $\Omega$.  This raises the question of whether the method is actually measuring $\Omega$.  To explore this possibility, a test was performed by applying the solution method to Holmberg II and IC 2574.  Both are dwarf irregular galaxies (RC3) that have no coherent or well organized patterns, and show material clumped around expanding H{\hskip 1.5pt \footnotesize  I} bubbles (Puche et al. 1992; Walter \& Brinks 1999).  Results for their pattern speeds would represent the averaged angular rotation of the individual clumps of material, and are therefore expected to differ somewhat from $\Omega$.

Data maps for both galaxies were obtained from the THINGS H{\hskip 1.5pt \footnotesize  I} galaxy survey.  Information about the observations, data reduction, and parameters of the maps can be found in \cite{w08}.  The rotation curves, as well as the position and orientation parameters, were adopted from those determined by \cite{o10} for Holmberg II and by S. -H. Oh et al. (2008, 2010, submitted) for IC 2574.  The position and orientation parameters are shown in Table 7.  The rotation curves, as well as the uncertainties for the parameters shown in Table 7, were provided by S. -H. Oh (2010, private communication).  In the rest of this section the descriptions of the data, the results, and the analysis are presented by galaxy.  These subsections are followed by conclusions of the test.

%t7
\begin{deluxetable*}{lcc}
\tablecaption{Adopted Parameters of Holmberg II and IC 2574}
\tablewidth{0pt}
\startdata
\tableline\tableline\\[-7.5 pt]
Parameter&Holmberg II& IC 2574\\
\tableline\\[-7.5 pt]
Kinematic center R. A. (J2000) &08$^{h}$ 34$^{m}$ 06.5$^{s}$ $\pm$ 8.25$\arcsec$& 10$^{h}$ 28$^{m}$ 27.7$^{s}$ $\pm$ 7.5$\arcsec$\\
Kinematic center DEC. (J2000) & \hskip 10.5pt  $+$70$\degr$ 43$\arcmin$ 24$\arcsec$ $\pm$ 12.45$\arcsec$&\hskip 2.6pt $+$68$\degr$ 24$\arcmin$ 59$\arcsec$ $\pm$ 15$\arcsec$\\
Inclination angle & \hskip 0pt 49$\degr$ $\pm$ 6$\degr$ & 55$\degr$ $\pm$ 6$\degr$\\
Position angle & 275$\degr$ $\pm$ 10$\degr$&53$\degr$ $\pm$ 2$\degr$\\
V$_{sys}$  & 156.5 $\pm$ 1 km s$^{-1}$&53 $\pm$ 1 km s$^{-1}$
\enddata
\end{deluxetable*}

\subsection{Holmberg II}%6.1

%%f11
\begin{figure}
\centering
\includegraphics[width=0.47\textwidth]{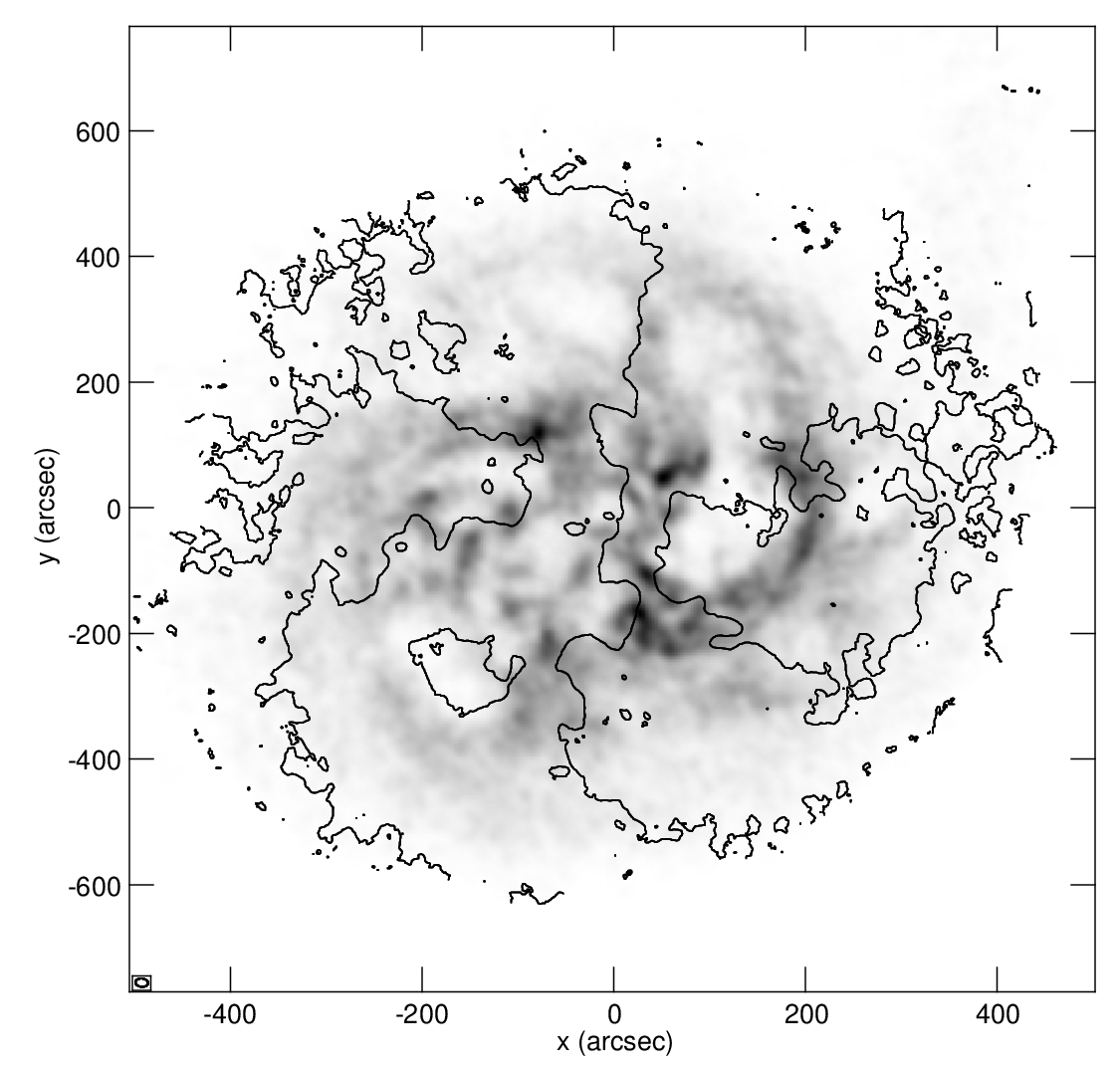} 
\caption{Rotated H{\hskip 1.5pt \footnotesize  I} specific intensity map of Holmberg II overlaid with contours of constant $v_y$.  The galaxy is oriented in the same way as shown in Figure 1.  The central, most vertical contour is $v_y$ = 0 km s$^{-1}$.  Velocity contours are spaced 26 km s$^{-1}$ apart.  The receding half of the galaxy is to the right.  The peak flux is 599 Jy m s$^{-1}$ beam$^{-1}$.  In the bottom left corner is a plot of the synthesized beam.} 
\end{figure}

Figure 11 shows the centered and rotated specific intensity map of Holmberg II overlaid with contours of constant $v_y$.  In the figure one can see the clumpy distribution of the H{\hskip 1.5pt \footnotesize  I}, and the lack of a coherent or well organized pattern.  Calculations of Equations (4) and (5) were restricted to the region $|y|$ $\leqslant$ 400$\arcsec$.  Different $y_j$ were spaced 9 pixels apart, which amounts to a separation distance that is 0.37$\arcsec$ larger than the FWHM of the synthesized beam in that direction.  This provided 36 independent calculations of Equations (4) and (5).  Consistent results were obtained using different regions of implementation and larger spacings between different $y_j$.

%%t8
\begin{deluxetable}{lccccc}
\tablecaption{Statistics for Holmberg II} 
\tablewidth{0pt}
\startdata
\tableline\tableline\\[-7.5 pt]
&&&&$P$&$R^2$ \\
Equation&$n$&$\chi_\nu^{2}$&$F$&($\%$)&($\%$)\\
 \tableline\\[-7.5 pt]
(2), (3)& 0 &  2.85 & \nodata & \nodata & \nodata  \\[2 pt]
\tableline\\[-7.5 pt]
(2) & 1 & 1.84  &  20.1 & $<$ 0.01  & 12.7 \\
& 2 & 1.87  &  \nodata  & \nodata  & 12.7 \\[2 pt]
\tableline\\[-7.5 pt]
(3) &1 &  2.03 & 15.2 & $<$ 0.01  & \hskip 4.5pt9.8 \\[2 pt]
& 2 & 2.09 & \nodata & \nodata & 10.0
\enddata
\end{deluxetable} 

The statistics for the results are shown in Table 8.  For both forms of $\Omega_p$, the improvement in $\chi_\nu^{2}$ is extremely statistically significant to order $n$ = 1 ($P$ $<$ 0.01\%).  Additional terms did not provide an improvement in $\chi_\nu^{2}$.  The $n$ = 1 solution using Equation (2) has the largest value of $F$ and $R^2$, and is therefore adopted as the best estimate of $\Omega_p$.  For this solution $\alpha_0$ = 0.29 ($\pm$ 0.12) km s$^{-1}$ arcsec$^{-1}$ and $\alpha_1$ = $-$ 3.16 ($\pm$ 3.76) $\times$ 10$^{-4}$ km s$^{-1}$ arcsec$^{-2}$.  

%f12
\begin{figure*}
\centering
\includegraphics[width=0.94\textwidth]{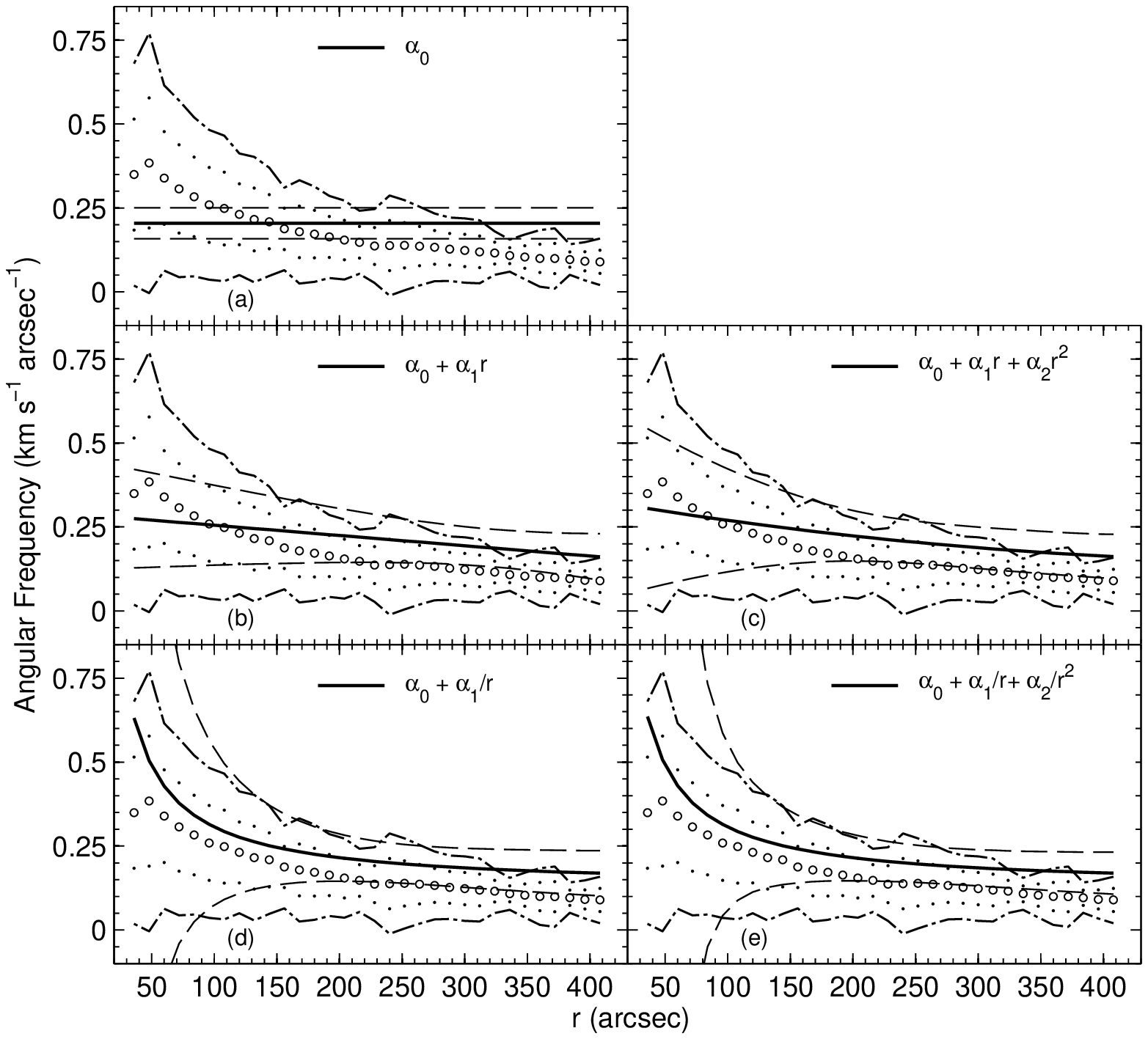} 
\caption{Plots of pattern speeds from the solutions for Holmberg II.  The figure is formatted the same way as Figure 5.  Panel (a) shows the mean pattern speed.  Panels (b), and (c) show pattern speeds having the form of Equation (2) to order $n$ = 1, and 2 respectively. Panels (d) and (e) show pattern speeds having the form of Equation (3) to order $n$ = 1 and 2 respectively.} 
\end{figure*}

Figure 12 shows plots of $\Omega_p$ for the solutions whose statistics are shown in Table 8.  The best estimate of $\Omega_p$ is shown in panel (b).  Compared with the sloping line in panel (b), the result in panel (a) is clearly the mean pattern speed.  There are more negative residuals in all of the panels.  All of the panels also show the pattern corotating with the material in the inner part of the disk, and rotating faster than the material in the outer disk.  There may be a faster rotating clump of material in the outer disk that is leveraging the solution.  If the method were measuring $\Omega$, then the result in panel (a) would be smaller, and the slope of the line in panel (b) would be larger.  

%f13
\begin{figure}
\centering
\includegraphics[width=0.47\textwidth]{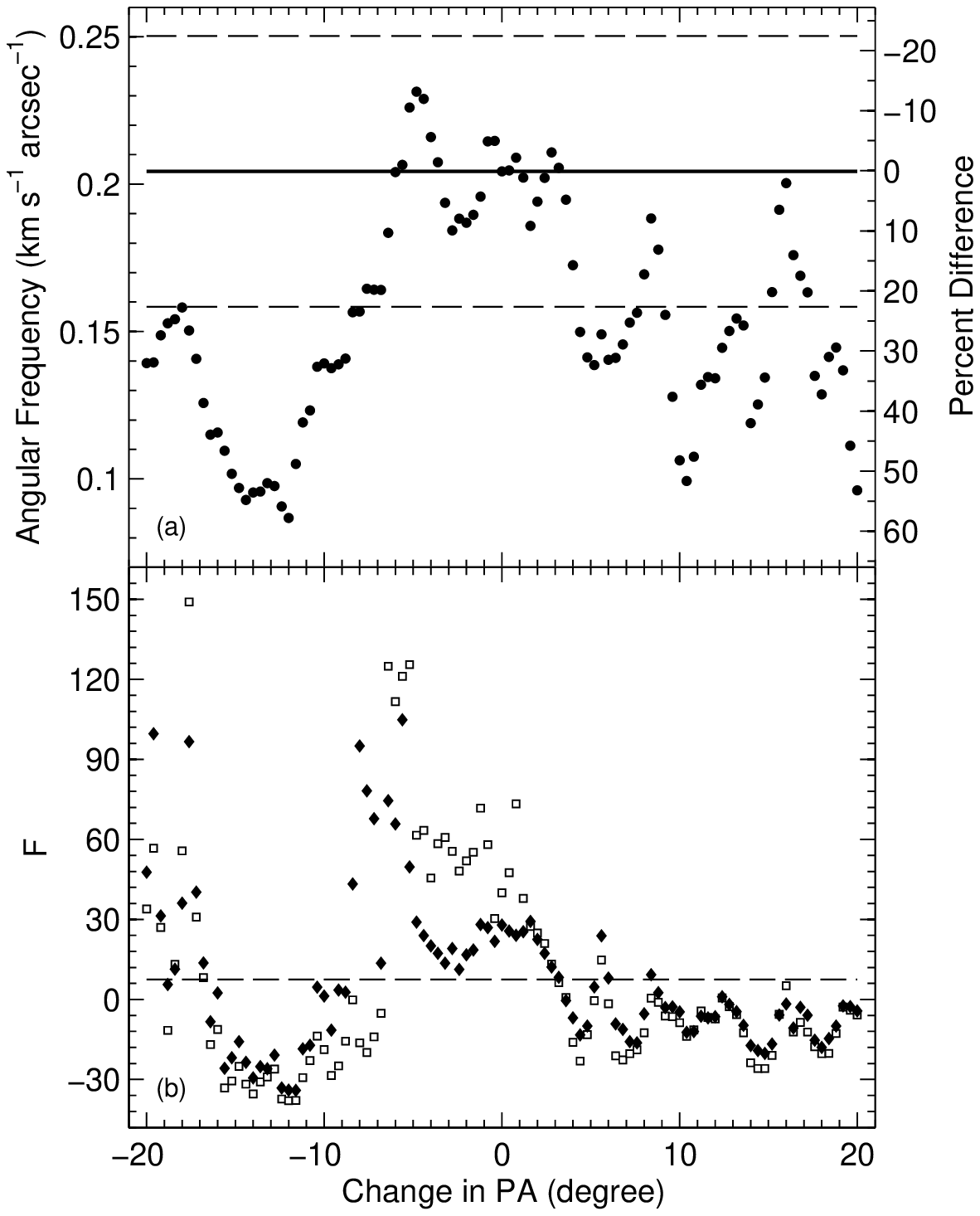} 
\caption{Effect of an incorrect PA for Holmberg II.  The figure is formatted the same way as Figure 6.} 
\end{figure}

The effect of an incorrect PA was explored in the same way as for NGC 1365.  The results are shown in Figure 13.  Most of the results for the mean pattern speed are smaller than the value obtained using the assumed PA.  The mean percent difference within $\pm$ 1$\sigma_{PA}$ is 13\%.  The largest percent difference within $\pm$ 1$\sigma_{PA}$ is $\sim$ 48\%.  The pattern speed for the solutions to order $n$ = 1 would be coincident with $\Omega$ for a percent difference $\sim$ 30\%, but in panel (b) one can see that these solutions generally do not provide a statistically significant improvement to $\chi_{\nu}^{2}$.  

\subsection{IC 2574}%6.2

%f14
\begin{figure}
\centering
\includegraphics[width=0.47\textwidth]{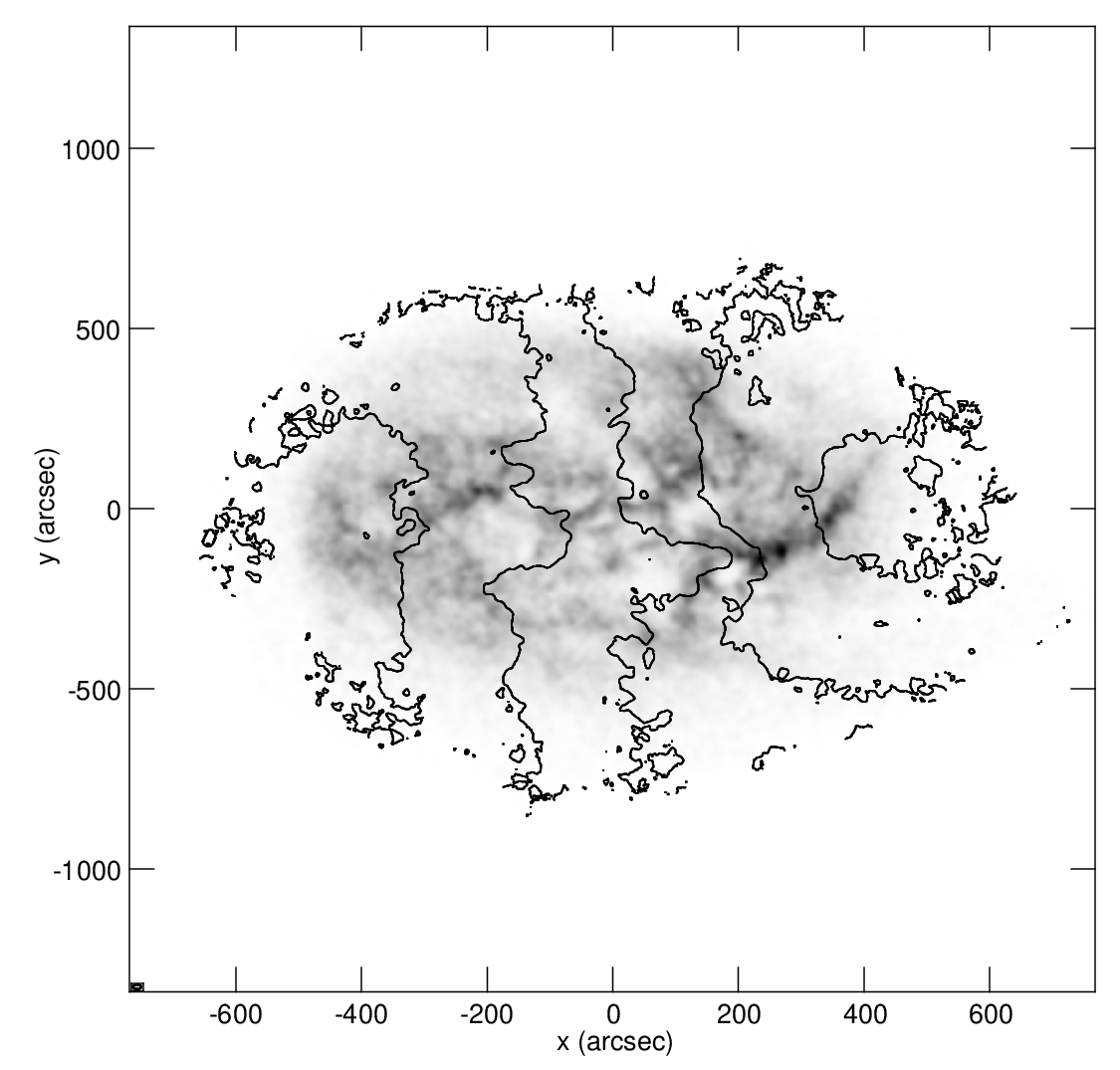} 
\caption{Rotated H{\hskip 1.5pt \footnotesize  I} specific intensity map of IC 2574 overlaid with contours of constant $v_y$.  The galaxy is oriented in the same way as shown in Figure 1.  The central, most vertical contour is $v_y$ = 0 km s$^{-1}$.  Velocity contours are spaced 28 km s$^{-1}$ apart.  The receding half of the galaxy is to the right.   The peak flux is 639 Jy m s$^{-1}$ beam$^{-1}$.  In the bottom left corner is a plot of the synthesized beam.} 
\end{figure}

Figure 14 shows the centered and rotated  specific intensity map of IC 2574 overlaid with contours of constant $v_y$.  Like Figure 11, one can see the clumpy distribution of the H{\hskip 1.5pt \footnotesize  I} and the lack of a coherent or well organized pattern.  Calculations of Equations (4) and (5) were restricted to the region $|y|$ $\leqslant$ 600$\arcsec$.  Different $y_j$ were spaced 9 pixels apart, which amounts to a separation distance that is 1.11$\arcsec$ larger than the FWHM of the synthesized beam in that direction.  This provided 52 independent calculations of Equations (4) and (5).  Consistent results were obtained using different regions of implementation and larger spacings between different $y_j$.

%t9
\begin{deluxetable}{lccccc}
\tablecaption{Statistics for IC 2574} 
\tablewidth{0pt}
\startdata
\tableline\tableline\\[-7.5 pt]
&&&&$P$&$R^2$ \\
Equation&$n$&$\chi_\nu^{2}$&$F$&($\%$)&($\%$)\\
 \tableline\\[-7.5 pt]
(2), (3)& 0 &  1.83  & \nodata & \nodata & \nodata  \\[2 pt]
\tableline\\[-7.5 pt]
(2)&1 & 2.82  & \nodata & \nodata & 7.0 \\[2 pt]
\tableline\\[-7.5 pt]
(3) &1 & 1.92 & \nodata & \nodata & 1.2
\enddata
\end{deluxetable} 

The statistics for the results are shown in Table 9.  For both forms of $\Omega_p$, the $n$ = 1 term did not provide an improvement in $\chi_\nu^{2}$.  The solution when $n$ = 0 is therefore adopted as the best estimate of $\Omega_p$.  For this solution $\alpha_0$ = 0.15 ($\pm$ 0.05) km s$^{-1}$ arcsec$^{-1}$.  The lack of convincing evidence for a radial dependence in $\Omega_p$ does not require that the pattern is rigidly rotating.  The solution method, or the simple functional forms used for $\Omega_p$, may be inadequate for detecting variations in the pattern speed of IC 2574.  

%f15
\begin{figure*}
\centering
\includegraphics[width=0.97\textwidth]{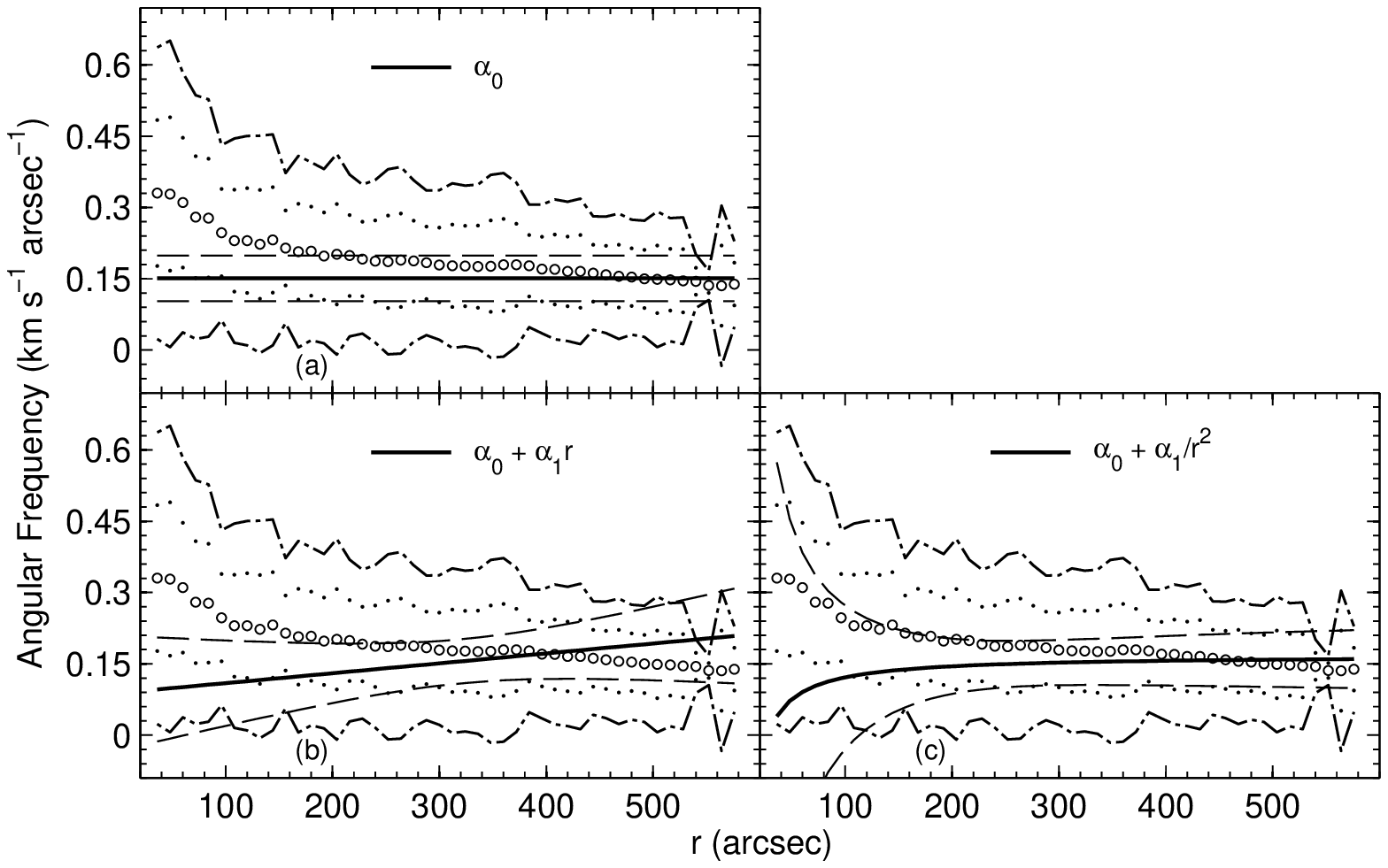} 
\caption{Plots of pattern speeds from the solutions for IC 2574.  The figure is formatted the same way as Figure 5.  Panel (a) shows the solution when $n$ = 0.  Panel (b) shows the pattern speed having the form of Equation (2) to order $n$ = 1.  Panel (c) shows the pattern speed having the form of Equation (3) to order $n$ = 1.} 
\end{figure*}

Figure 15 shows plots of $\Omega_p$ for the solutions whose statistics are shown in Table 9.  Like the results for Holmberg II, none of the pattern speeds shown resembles $\Omega$.   There is a clear downward trend in $\Omega$ that is not seen in any of the plots of $\Omega_p$.  Panel (a) shows the best estimate of $\Omega_p$, which is less than $\Omega$ for most of the galaxy.  If the method is measuring $\Omega$, then the result in panel (a) would be larger.  All three panels show the pattern corotating with the material in the outer part of the disk, specifically around 500$\arcsec$.  If the method is measuring $\Omega$, then the result in panel (a) would be larger. 

%f16
\begin{figure}
\centering
\includegraphics[width=0.47\textwidth]{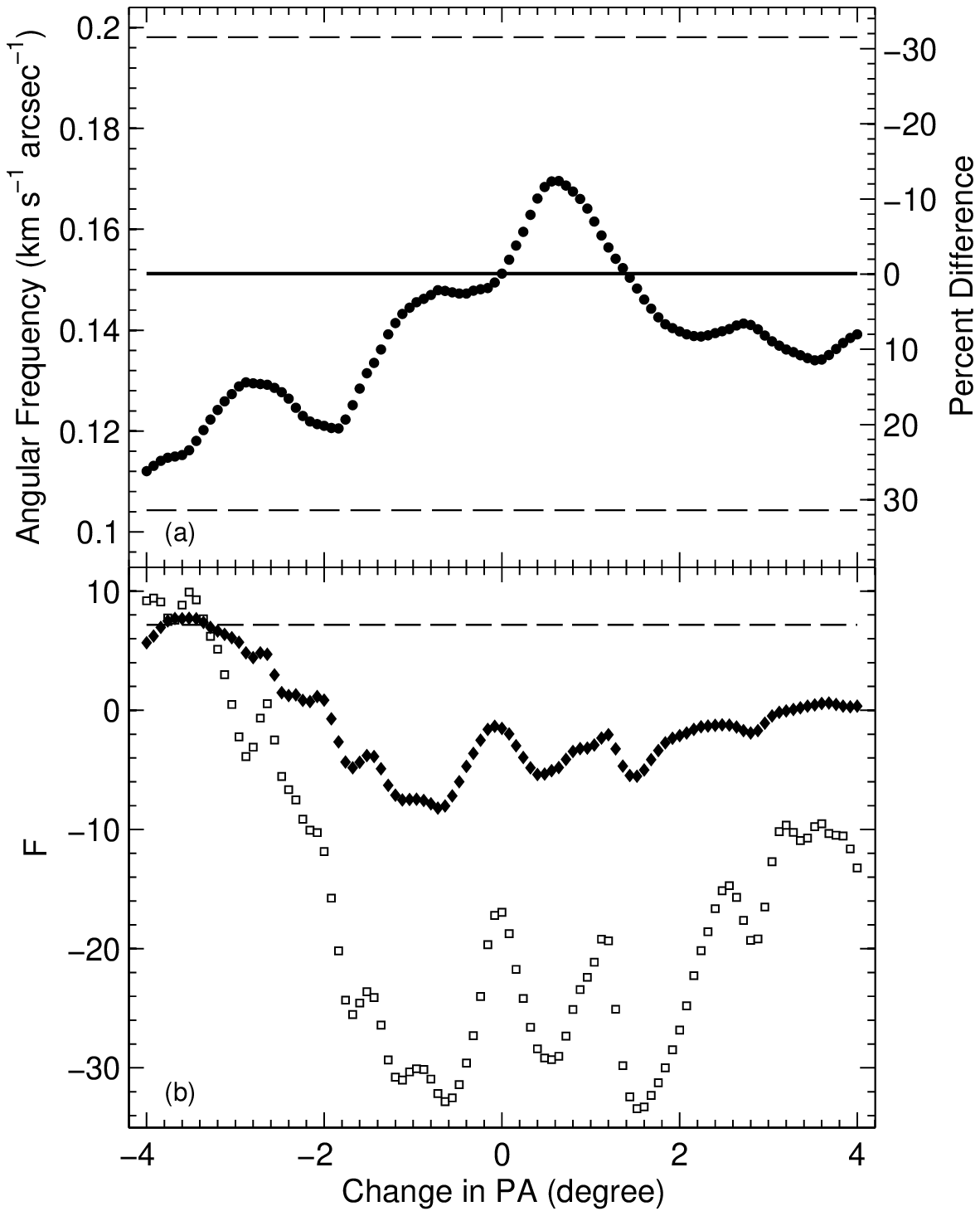} 
\caption{Effect of an incorrect PA for IC 2574.  The figure is formatted the same way as Figure 6.} 
\end{figure}

The effect of an incorrect PA was explored in the same way as for NGC 1365.  The results are shown in Figure 16.  All of the results for the mean pattern speed fit within the 95\% confidence interval for the value obtained using the assumed PA.  The mean percent difference within $\pm$ 1$\sigma_{PA}$ is 2\%.  The largest percent difference within $\pm$ 1$\sigma_{PA}$ is $\sim$ 22\%.  For all of $\pm$ 1$\sigma_{PA}$, and most of $\pm$ 2$\sigma_{PA}$, the $n$ = 1 solutions did not provide a statistically significant improvement in $\chi_\nu^{2}$.  The largest pattern speed shown in panel (a), $\sim$ 0.17 km s$^{-1}$ arcsec$^{-1}$, would not provide much of a better fit to the material speed than the value obtained using the assumed PA.

\subsection{Conclusions}%6.3

The results for Holmberg II and IC 2574 are two examples of when $\Omega_p$ is not very similar to $\Omega$.  Unlike plots of the results for NGC 1365, plots of the results for Holmberg II and IC 2574  suggest specific regions of corotation.  Their results are consistent with patterns that are decoupled from the material.  If the spiral pattern of NGC 1365 is decoupled from the material, the difference may be too small to distinguish with the precision of the results.  It is also possible that the assumed PA of NGC 1365 is incorrect, in which case the pattern is most likely rotating more slowly than the material.  

%t10
\begin{deluxetable*}{lll}
\tablecaption{Results for Holmberg II and IC 2574 in Common Units} 
\tablewidth{0pt}
\startdata
\tableline\tableline\\[-7.5 pt]
Galaxy&\multicolumn{2}{c}{Result}\\
\tableline\\[-7.5 pt]
Holmberg II&Mean pattern speed&\hskip 0.8ex12.64 ($\pm$ 4.02) km s$^{-1}$ kpc$^{-1}$\\
&Mean shear rate& $-$1.17 ($\pm$ 0.80) km s$^{-1}$ kpc$^{-2}$\\
&Best estimate of $\Omega_p$& \hskip 0.8ex17.7 ($\pm$ 6.47) km s$^{-1}$ kpc$^{-1}$ \\
&&\hskip 3.6ex  $-$1.17 ($\pm$ 0.80) km s$^{-1}$ kpc$^{-2}$ $r$ \\[2pt]
\tableline\\[-7.5 pt]
IC 2574&Best estimate of $\Omega_p$&  \hskip 0.8ex8.10 ($\pm$ 1.92) km s$^{-1}$ kpc$^{-1}$
\enddata
\end{deluxetable*} 

The most important results for Holmberg II and IC 2574 are summarized in Table 10 using common units.  The units were converted using an assumed distance of 3.33 ($\pm$ 0.99) Mpc to Holmberg II and 3.85 ($\pm$ 0.68) Mpc to IC 2574.  The distances and their uncertainties were estimated from the mean of 6 redshift-independent measurements for Holmberg II, and 11 such measurements for IC 2574, all found in the NED. 

\section{COMPARISON WITH THE RESULTS FROM OTHER SOLUTION METHODS}%7

\subsection{The Original Solution Method of TW84}%7.1

TW84 originally sought a solution for bar pattern speeds, allowing them to pull $\Omega_p$ outside the integral in Equation (1).  Carrying through the integration in $x$ for the second term on the left-hand side of Equation (1) removes that term because of the boundedness of $I$.  Subsequent integration from $y$ to +$\infty$, or similarly for negative $y$, removes the spatial derivatives.  Then, Equation (1) reduces to,
%e21
\begin{equation}
\Omega_p\int_{-\infty}^{+\infty}I(x,y)\,x\,dx = \int_{-\infty}^{+\infty}I(x,y)\,v_y(x,y)\,dx.
\end{equation}
Several independent calculations at different distances from the kinematic major axis can be used to find an average of the results and measure the uncertainty.  It is common practice to normalize Equation (21) by $\int_{-\infty}^{+\infty}I(x,y)\,dx$ \cite{mk95}.

If the spiral arms rotate with a constant $\Omega_p$, then the original solution method is applicable to those patterns as well.  However, there are many instances where that the results suggest otherwise.  Results for NGC 3031 (W98), NGC 2915\cite{b99}, NGC 5194\cite{z04}, NGC 1068\cite{rw04},  NGC 4321\cite{h05}, and NGC 6946\cite{f07} show a decreasing trend for increasing $|y|$.  

Such trends are consistent with a shearing pattern.  Consider that according to the mean value theorem for integration,
%e22
\begin{eqnarray}
&&\int_{y}^{+\infty}\int_{-\infty}^{+\infty}\Omega_p(r)\Big\{y{\partial \over \partial x} I(x,y)- x{\partial \over \partial y}I(x,y)\Big\}\,dx\,dy \nonumber \\ &&\hskip 90pt = \overline{\Omega}_p(\bar{r})\int_{-\infty}^{+\infty}I(x,y_j)\,x\,dx,
\end{eqnarray}
where $\overline{\Omega}_p(\bar{r})$ is the mean pattern speed for the region of integration, and $\bar{r}$ is located somewhere in the region\cite{gr07}.  The lower limit for values of $\bar{r}$ increases with increasing $|y|$.  Equation (22) shows  that if $\Omega_p$ is decreasing with radius, then the result from calculating Equation (21) will be $\overline{\Omega}_p(\bar{r})$, whose value should show a decreasing trend with increasing $|y|$.

The original solution method of TW84 was applied to the H{\hskip 1.5pt \footnotesize  I} spiral pattern of NGC 1365 to see if the same statistics explained in Section 3 would show evidence for a decreasing trend in the results plotted versus $|y|$.  Individual calculations of Equation (21) were spaced 5 pixels apart, and covered the same region as was used for the current method ($|y|$ $\leqslant$ 400$\arcsec$).  The half-widths of the 95\% confidence intervals for the coefficients, and 95\% confidence bands for the trend lines fit to the results, were calculated in a similar way as Equations (17) and (18), with the exception that the standard errors were estimated from the covariance matrix.  

%f17
\begin{figure*}
\centering
\includegraphics[width=0.94\textwidth]{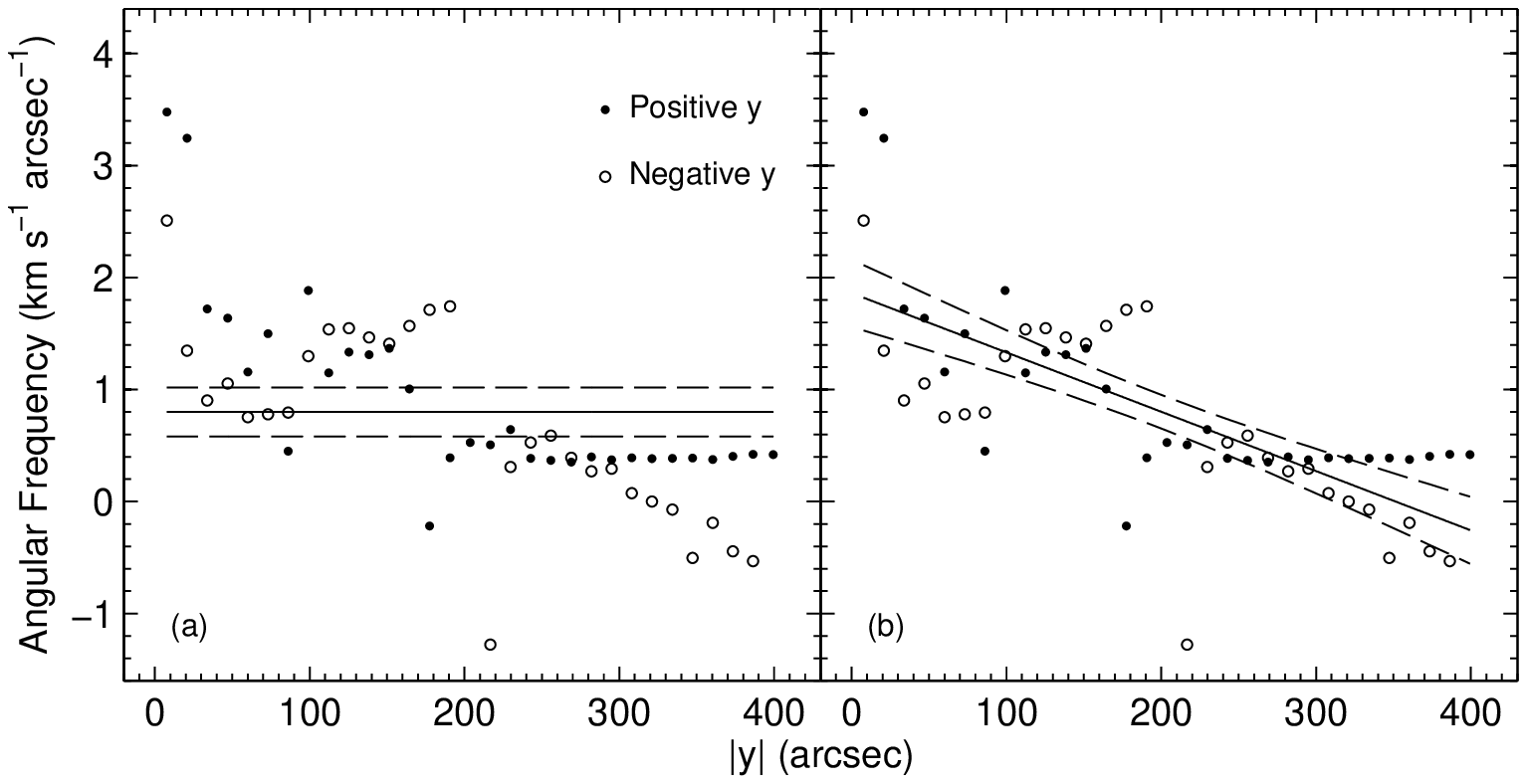} 
\caption{Plots of trend lines fit to the results from applying the original TW84 solution method to NGC 1365.  The points and open circles represent calculations of Equation (21) for positive and negative $y$ respectively.  The solid lines are trends fit to the results.  The dashed lines are 95\% confidence bands.  In panel (a) is a constant trend line.  In panel (b) is a sloping trend line.} 
\end{figure*}

Figure 17 shows constant and sloping trend lines fit to the results.  The constant trend line at 0.80 ($\pm$ 0.22) km s$^{-1}$ arcsec$^{-1}$ in panel (a) is less than the mean pattern speed of 1.27 ($\pm$ 0.10) km s$^{-1}$ arcsec$^{-1}$ obtained using the current method.  The difference is due to the lower value of $\Omega_p$ in the outer disk contributing to $\overline{\Omega}_p(\bar{r})$ for all calculations.  Note that in panel (a) there are more positive residuals for smaller $|y|$ and more negative residuals for larger $|y|$, which occurs separately for both positive and negative $y$. The sloping trend line in panel (b) has an intercept on the vertical axis at 1.86 ($\pm$ 0.30) km s$^{-1}$ arcsec$^{-1}$ and a slope of $-$ 0.53 ($\pm$ 0.13) $\times$ 10$^{-2}$ km s$^{-1}$ arcsec$^{-2}$.  An $F$ test showed that the sloping trend line provides a statistically significant improvement to $\chi_\nu^2$ ($P$ = 1.32\%).  

\subsection{Allowing for $\Omega_p$ to Vary With Radius}%7.2

\subsubsection{The Solution Method of E94}%7.2.1

When $\Omega_p$ is allowed to vary with radius, solutions of the continuity equation integrated over $x$ and $y$ are non-trivial.  Such solutions were first explored by E94, who showed that a numerical solution is possible by making a switch to polar coordinates.  The result is a Volterra equation of the first kind for $\Omega_p$,
%e23
\begin{equation}
\int_{r=y_j}^{+\infty}\Omega_p(r)\mbox{K}_{j,i}(r_i,y_j)\,dr= \int_{-\infty}^{+\infty}I(x,y_j)\,v_y(x,y_j)\,dx,
\end{equation}
where,
%e24
\begin{equation}
\mbox{K}_{j,i}=r_i\Big\{I(\sqrt{r_i^{2}-y^{2}_j},y_j)-I(-\sqrt{r_i^{2}-y^{2}_j},y_j)\Big\},
\end{equation}
or similarly for negative $y$.  For an illustration showing the radial binning in Equation (18), see Figure 1 of Meidt et al. 2008a.  Equation (23) is solvable using back substitution methods, but as E94 points out, such methods are ill-advised because of the triangular shape of {\bf K} and its sensitivity to symmetries in the data.  This was indeed found to be the case by M06, whose solutions for the CO spiral pattern of NGC 1068 became increasingly unstable for smaller radii.  They showed examples of some of the results for large radial binnings in the discretization of Equation (23) that produced a declining trend in $\Omega_p$ with radius.  

The method of E94 was applied to the H{\hskip 1.5pt \footnotesize  I} spiral pattern of NGC 1365 to see how the results compared with those obtained using the current method, but the solutions were very unstable.  Radial binnings as small as the synthesized beam produced results that oscillated by up to 4 orders of magnitude.  The results for larger binnings ranging from 100$\arcsec$ - 200$\arcsec$ oscillated by an order of magnitude.  The wildly oscillating behavior and the lack of a consistent trend make these results unreliable, and are therefore not reported.  A single radial binning from 50$\arcsec$ to 400$\arcsec$, however, produced a mean pattern speed of 1.14 ($\pm$ 0.06) km s$^{-1}$ arcsec$^{-1}$ for negative $y$, and 1.36 ($\pm$ 0.07) km s$^{-1}$ arcsec$^{-1}$ for positive $y$, in agreement with the results for regions $\mathcal{C}$ and $\mathcal{D}$ respectively in Table 5.  The uncertainties reported for these solutions may be larger because they do not include an estimate of the error introduced by rebinning the data into cylindrical coordinates.  The average of the two results is 1.25 ($\pm$ 0.09) km s$^{-1}$ arcsec$^{-1}$, in agreement with the result of 1.27 ($\pm$ 0.10) km s$^{-1}$ arcsec$^{-1}$ obtained using the current method.

Regularization can be used to obtain stable solutions of Equation (23), but such solutions are not desirable for determining the radial behavior of $\Omega_p$.  Regularization makes standard tools such as $\chi^{2}$ statistics inappropriate for calculating confidence intervals or evaluating the goodness of fit \cite{a05}.  This is because regularization biases the solution and this bias is often much larger than the confidence intervals that would be obtained.  Regularization has been tried by Meidt et al. (2008b) and M09, but any conclusions based on $\chi^{2}$ statistics are unreliable. 

%\vskip 15pt
\subsubsection{The Solution Method of W98}%7.2.2

When W98 measured the pattern speed of NGC 3031, four different, albeit similar, analytic solutions for a constant $\Omega_p$ were derived.  Only one such solution is shown here to demonstrate the method.  It consists of integrating Equation (21) once from $y$ = $y_1$ to +$\infty$, and then again from $y$ = $y_2$ $>$ $y_1$ to +$\infty$, or similarly for negative $y$.  The spacing of $y_1$ and $y_2$ should be more than a synthesized beam apart to insure that the calculations are independent.  After subtracting the two integrations, the solution is,
%e25
 \makeatletter 
 \def\@eqnnum{{\normalsize \normalcolor (\theequation)}} 
  \makeatother
  {\small
\begin{equation}
\Omega_p = {{ \int_{-\infty}^{+\infty}\!I_\nu(x,y_1)\,v_y(x,y_1)\,dx -  \int_{-\infty}^{+\infty}\!I_\nu(x,y_2)\,v_y(x,y_2)\,dx} \over { \int_{-\infty}^{+\infty}\!I_\nu(x,y_1)\,x\,dx - \int_{-\infty}^{+\infty}\!I_\nu(x,y_2)\,x\,dx}}\,,
\end{equation}}
and similarly for negative $y$.  This solution is equivalent to integrating the continuity equation over the region of a strip having a width of $|y_2|$ $-$ $|y_1|$.  

The region of integration in Equation (25) can be much smaller than the region in Equation (21).  This makes it possible to better constrain the coordinate $\bar{r}$ in the mean value theorem for integration.  For each calculation, W98 estimated $\bar{r}$ as the intensity weighted mean radius in a strip.  Constant and sloping trend lines were then fit to the results plotted versus $\bar{r}$.  The sloping trend line provided a statistically significant improvement to $\chi_\nu^2$, and a measure of the mean shear rate for the spiral pattern of NGC 3031.

The method of W98 was applied to the H{\hskip 1.5pt \footnotesize  I} spiral pattern of NGC 1365 to see how the results compare with those obtained using the current method.  It was applied over the same region as the current method ($|y|$ $\leqslant$ 400$\arcsec$), and used a spacing for $y_1$ and $y_2$ of 5 pixels.  Individual calculations of Equation (25) were spaced 5 pixels apart.  

%f18
\begin{figure*}
\centering
\includegraphics[width=0.94\textwidth]{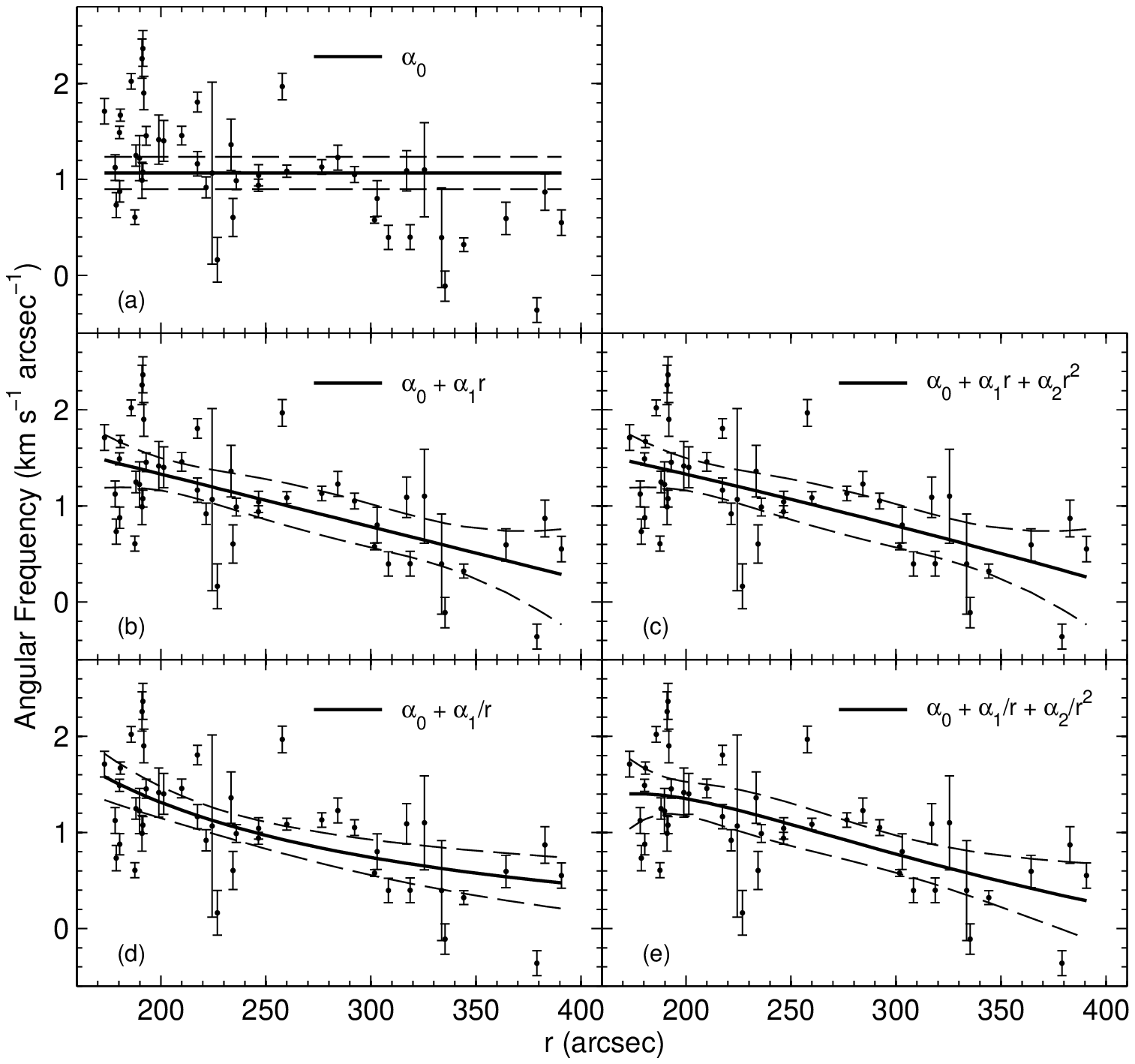} 
\caption{Plots of pattern speeds fit to the results from applying the solution method of W98 to NGC 1365.  The points represent calculations of Equation (25).  The error bars represent the uncertainties propagated through the calculations.  The solid lines are pattern speeds fit to the results.  The dashed lines are 95\% confidence bands.  Panel (a) shows the mean pattern speed.  Panels (b) and (c) show pattern speeds having the form of Equation (2) to order $n$ = 1 and 2 respectively. Panels (d) and (e) show pattern speeds having the form of Equation (3) to order $n$ = 1 and 2 respectively.} 
\end{figure*}

Figure 18 shows the fits that were made to the results using Equations (2) and (3) as forms of $\Omega_p$.  Due to the nature of $\bar{r}$, there are no results inside of 170$\arcsec$.  The uncertainties propagated through the calculations of $\bar{r}$ were negligible, therefore the standard errors were found using the covariance matrix.  There is hardly any noticeable difference between panels (b) and (c) of Figure 18.  The pattern speed in panel (e) appears to be over-fitting the results. 

%t11
\begin{deluxetable*}{lcccc}
\tablecaption{Best-Fit Coefficients for NGC 1365 Using the Method of W98} 
\tablewidth{0pt}
\startdata
\tableline\tableline\\[-7.5 pt]
&&$\alpha_0$ $\pm$ HW$_0$&$\alpha_1$ $\pm$ HW$_1$&$\alpha_2$ $\pm$ HW$_2$ \\
Equation&$n$&(km s$^{-1}$ arcsec$^{-1}$)&10$^{-2}$ (km s$^{-1}$ arcsec$^{-2}$)&10$^{-4}$ (km s$^{-1}$ arcsec$^{-3}$)\\
 \tableline\\[-7.5 pt]
(2), (3) & 0 &  1.07 $\pm$ 0.17 & \nodata & \nodata\\[2 pt]
\tableline\\[-7.5 pt]
(2) &1 &  2.43 $\pm$ 0.53 &$-$0.55 $\pm$ 0.21 & \nodata \\
&2 &  2.25	 $\pm$ 2.66 &$-$0.40 $\pm$ 2.07  & $-$0.03 $\pm$ 0.38\\ [2 pt]
\tableline\\[-7.5 pt]
& & 10$^{-1}$ (km s$^{-1}$ arcsec$^{-1}$)  & 10$^{2}$ (km s$^{-1}$) & 10$^{3}$ (km s$^{-1}$ arcsec) \\[2 pt]
\cline{3-5}\\[-7.5 pt]
(3) &1 & $-$4.05 $\pm$ 6.00 & 3.44 $\pm$ 1.36 & \nodata\\
& 2 & $-$23.1 $\pm$ 29.5 & 13.1 $\pm$ 14.7 & $-$116 $\pm$ 176  
\enddata
\end{deluxetable*} 

%
%\begin{deluxetable*}{lccccc}
%\tablecaption{Best-Fit Coefficients for NGC 1365} 
%\tablewidth{0pt}
%\startdata
%\tableline\tableline\\[-7.5 pt]
%& &$\alpha_0$ $\pm$ HW$_0$&$\alpha_1$ $\pm$ HW$_1$&$\alpha_2$ $\pm$ HW$_2$&$\alpha_3$ $\pm$ HW$_3$ \\
%Equation &$n$&(km s$^{-1}$ arcsec$^{-1}$) &10$^{-2}$ (km s$^{-1}$ arcsec$^{-2}$)&10$^{-4}$ (km s$^{-1}$ arcsec$^{-3}$)&10$^{-7}$ (km s$^{-1}$ arcsec$^{-4}$)\\
%\tableline\\[-7.5 pt]
%(2), (3) & 0 &  1.27 $\pm$ 0.10 & \nodata & \nodata & \nodata \\[2 pt]
%\tableline\\[-7.5 pt]
%(2) &1 &  2.68 $\pm$ 0.38 &$-$0.55 $\pm$ 0.13 & \nodata & \nodata \\
%&2 &  4.05	 $\pm$ 0.83 &$-$1.65 $\pm$ 0.57  & 0.20 $\pm$ 0.09 & \nodata \\  
%&3 &  5.93 $\pm$ 1.43 & $-$4.06 $\pm$ 1.64 & 1.11 $\pm$ 0.59 & $-$1.05 $\pm$ 0.64 \\[2 pt]
%\tableline\\[-7.5 pt]
%& & 10$^{-1}$ (km s$^{-1}$ arcsec$^{-1}$)  & 10$^{2}$ (km s$^{-1}$) & 10$^{3}$ (km s$^{-1}$ arcsec) &  \nodata \\[2 pt]
%\cline{3-6}\\[-7.5 pt]
%(3) &1 & $-$0.59 $\pm$ 2.88 & 2.87 $\pm$ 0.61& \nodata & \nodata \\
%& 2 & $-$1.04 $\pm$ 4.28 & 3.03 $\pm$ 1.29 & 0.95 $\pm$ 6.65 & \nodata 
%\enddata
%\end{deluxetable*} 

The coefficients for the fitted pattern speeds are shown in Table 11.  The mean pattern speed and the  $i$ = 1 coefficient for the fit using Equation (3) to order $n$ = 1 are in agreement with the values obtained using the current method.  The mean shear rate is exactly the same as the value obtained using the current method.
 
 %t12
\begin{deluxetable}{lccccc}
\tablecaption{Statistics for NGC 1365 Using the Method of W98} 
\tablewidth{0pt}
\startdata
\tableline\tableline\\[-7.5 pt]&&&&$P$&$R^2$ \\
Equation&$n$&$\chi_\nu^{2}$&$F$&($\%$)&($\%$)\\
 \tableline\\[-7.5 pt]
(2), (3)& 0 &  25.5  &   \nodata & \nodata & \nodata  \\[2 pt]
\tableline\\[-7.5 pt]
(2) &1 & 13.0  &  45.6 & $<$ 0.01  & 38.2 \\
&2 &  13.3 & \nodata  &  \nodata & 38.3 \\ [2 pt]
\tableline\\[-7.5 pt]
(3) &1 &  3.5 & 289  & $<$ 0.01 & 36.1\\
&2 & 0.37  &  383  & $<$ 0.01 & 38.5
\enddata
\end{deluxetable}
 
The statistics for the fits are shown in Table 12.  The method of W98 also produces convincing evidence for shear in the pattern, and an approximate 1/$r$ behavior for $\Omega_p$.  The improvement in $\chi_\nu^2$ for the fit when $n$ = 1 is extremely statistically significant for both forms of $\Omega_p$ ($P$ $<$ 0.01\%). For fits using Equation (2), adding the $n$ = 2 term did not provide an improvement in $\chi_\nu^2$.  For fits using Equation (3), adding the $n$ = 2 term produced a very small value of $\chi_\nu^2$ (= 0.37), which is a sign of over-fitting the results.  The large values of $\chi^{2}_\nu$ for the other fits, and the small values of $R^2$ for all of the radially dependent fits, are due to the large amount of scatter seen in Figure 18. 

%f19
\begin{figure*}
\centering
\includegraphics[width=0.94\textwidth]{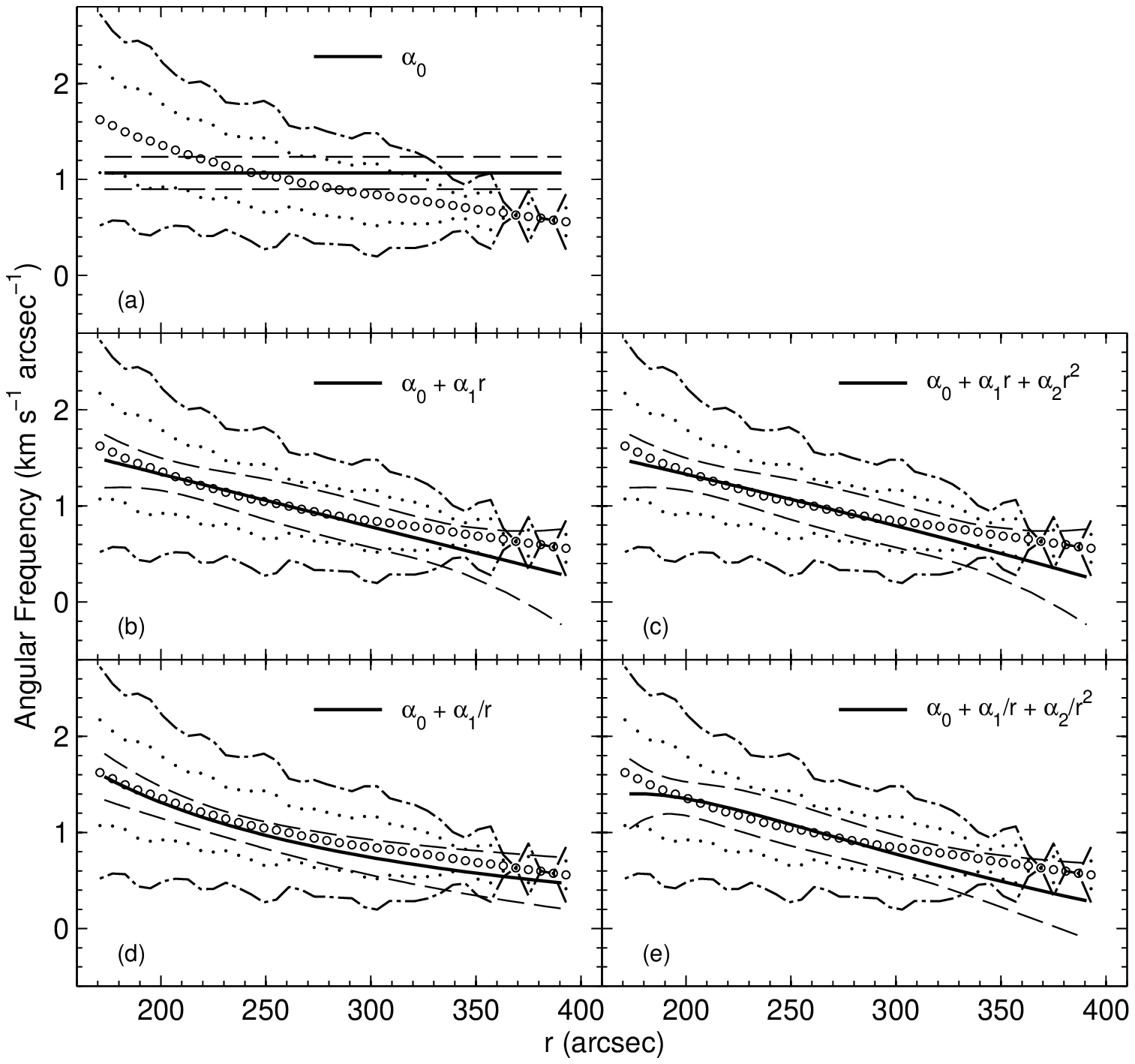} 
\caption{Plots of the results for NGC 1365 obtained using the solution method of W98.  The figure is formatted the same way as Figure 5.  Panel (a) shows the mean pattern speed.  Panels (b) and (c) show pattern speeds having the form of Equation (2) to order $n$ = 1 and 2 respectively. Panels (d) and (e) show pattern speeds having the form of Equation (3) to order $n$ = 1 and 2 respectively.} 
\end{figure*}

Figure 19 shows plots of the results that include possible locations for resonances.  When compared to the plots in Figure 5, they show how well the results agree with those obtained using the current method.  The plots in panels (a), (b) and (d) resemble those in panels (a), (b) and (e) respectively of Figure 5.  Note that in panel (d) of Figure 19, the fit using Equation (3) to order $n$ = 1 is very similar to $\Omega$.  There are also no clear indications of resonances associated with this solution.  The excellent agreement of the two different sets of results confirms the validity of the two different solution methods and the robustness of the conclusions arrived at for NGC 1365.  

\section{COMPARISON WITH PREVIOUS ESTIMATES OF $\Omega_p$}%8

The spiral arm pattern speed of NGC 1365 has been estimated previously using other methods that relied on the assumptions of the density-wave interpretation.  JvM95 estimated $\Omega_p$ by identifying possible signs of corotation and Lindblad resonances in the morphology and kinematics of NGC 1365.  P. A. Lindblad et al. (1996) estimated $\Omega_p$ by matching barred spiral galaxy models to observations.  The pattern speed of a model was set by an assumed corotation radius just outside the end of the bar.  Vera-Villamizar et al. (2001) estimated $\Omega_p$ by comparing infrared and blue images. They assumed that an azimuthal profile phase difference of zero for the two images is an indication of corotation.  
%t13
\begin{deluxetable*}{lcc}
\tablecaption{Previous Estimates of $\Omega_p$} 
\tablewidth{0pt}
\startdata
\tableline\tableline\\[-7.5 pt]
&$\Omega_p$ $\pm$ $\sigma$&Corotation Radius \\ 
 Reference&(km s$^{-1}$ arcsec$^{-1}$)&(arcsec)\\
 \tableline\\[-7.5 pt]
JvM95 & \hskip 0pt 2.1 \hskip 5pt $\pm$ 0.1 & 138 \\
P. A. B. Lindblad et al. (1996)  & \hskip 0pt  1.7 \hskip 5pt $\pm$ 0.1 & 157 \\ 
Vera-Villamizar et al. (2001) & \hskip 0pt 2.7 \hskip 5pt $\pm$ 0.2 & 113 \\
Average ................................... &  \hskip 0pt 2.2 \hskip 5pt $\pm$ 0.2 & 144 \\
\hskip 10pt This work (mean of $\Omega_p$) & \hskip 4.5pt 1.27 $\pm$ 0.05&\hskip 3pt\nodata
\enddata
\end{deluxetable*} 

The previous estimates of $\Omega_p$ and their corresponding radii of corotation resonances are summarized in Table 13.  Originally reported in common units, the previous estimates were converted to instrumental units using the distances to NGC 1365 that were adopted by the references in the table.  Also included in the table are the average of the previous estimates and the mean pattern speed obtained using the current method.  For consistency with the previous estimates, the uncertainty shown for the mean pattern speed is the 1$\sigma$ uncertainty, which is approximated as the standard error.  

The previous estimates are inconsistent with the results obtained using the current method.  They differ from the current result for the mean by at least 3$\sigma$, and their average is 73$\%$ larger than the current result for the mean. Furthermore, none of the corotation radii in Table 13 are clearly indicated in Figure 5.  Also note that none of the previous estimates are in agreement with each other.  They differ among themselves by at least 3$\sigma$.  

\section{RELEVANCE TO THEORIES OF SPIRAL STRUCTURE}%9

The results for NGC 1365 are inconsistent with fundamental assumptions in the density-wave interpretation of spiral structure that are often used for measuring spiral arm pattern speeds, i.e., a constant pattern speed with unique locations of corotation and Lindblad resonances.  They are more consistent with the sheared gravitational instabilities of Goldreich \& Lynden-Bell (1965), the swing amplifier of Toomre (1981), and the recurring transient patterns in the simulations described by Sellwood (2000, 2011).  The large amount of shear is especially problematic for applying the quasi-stationary spiral structure (QSSS) hypothesis of Lin \& Shu (1964, 1966), which proposes that grand-design spiral patterns are density waves that rotate approximately rigidly within a few dynamical timescales.  This hypothesis is appealing if one assumes that such patterns are long lived because it avoids the winding dilemma.  For a detailed discussion of spiral density waves, its successes in qualitatively explaining many observed properties of spiral galaxies, and the QSSS hypothesis, see Bertin \& Lin (1996) and Bertin (2000, Chapter 18).  

The spiral pattern of NGC 1365 is winding on a characteristic timescale of $\sim$ 500 Myrs (using the convention in M06).  A similar timescale was found by M06 for NGC 1068, and follows from the results of W98 for NGC 3031.  Winding is regularly seen in simulations of spiral galaxies.  Examples can be found in the simulations of isolated galaxies by Berman et al. (1978), Bottema (2003), and Dobbs \& Bonnell (2008); and of patterns generated from a tidal perturbation by Sorensen (1985), S. Oh (2008), Dobbs et al. (2010), and Struck et al. (2011, in press).  It is worth noting that the latter two authors provide plots of $\Omega_p$ for their simulated galaxies that show an approximately 1/$r$ behavior.

For such a short winding timescale the winding dilemma is avoidable if the spiral pattern is a recurring transient feature.  Support for the existence of transient spiral patterns can be found in the redistribution of angular momentum by spiral arm torqueing (Gnedin et al. 1995;  Binney \& Tremaine 2008, Section 6.1.5; Foyle et al. 2010), the damping of spiral arm amplitudes by viscous gas (Toomre 1969), and the inability of simulations to produce long-lived spiral patterns (Sellwood 2011).  In the case of NGC 1365, the bar could regenerate the spiral pattern (Sanders \& Huntley 1976; Huntley et al. 1978).  

A successful theory of spiral structure must be able to account for the shear observed in the spiral arms of NGC 1365.  A more detailed discussion of the consequences of the results for theories of spiral structure would be premature due to having only applied a rigorous analysis to the results for NGC 1365.  The new solution method should be applied to a sample of galaxies and the results thoroughly analyzed before it would be appropriate to do so.  A preliminary analysis of the results for NGC 2403, NGC 2903, NGC 3031, NGC 3627, NGC 4535, and NGC 5194, however, do show convincing evidence for shear in their spiral patterns as well.  These results, their thorough analysis, and additional tests of the method will be presented in forthcoming papers. 

\section{SUMMARY}%10

In this paper a new method was developed for solving the TW84 equations. It uses least squares to solve an integral equation for simple functional forms of $\Omega_p$.  Standard statistical tools can then be used to determine the functional form of $\Omega_p$ that provides the best solution, and thus the best estimate of $\Omega_p$.  The method was applied to the H{\hskip 1.5pt \footnotesize  I} spiral pattern of the barred grand-design galaxy NGC 1365.  The findings are as follows:

1.)  The mean pattern speed is 14.5 ($\pm$ 2.16) km s$^{-1}$ kpc$^{-1}$.  

2.)  The mean shear rate is $-$ 0.71 ($\pm$ 0.24) km s$^{-1}$ kpc$^{-2}$.  

3.)  The best estimate of $\Omega_p$ is 2.87 ($\pm$ 0.61) $\times$10$^{2}$ km s$^{-1}$ /$r$.  

4.)  The radial behavior of $\Omega_p$ is approximately 1/$r$. 

5.)  There are no clear indications of unique corotation or Lindblad resonances.

6.)  Tests of the method showed that these findings are not selection biased.  Consistent results were obtained for different regions of the galaxy.
    
7.)  The pattern speed is very similar to $\Omega$.  If the pattern is decoupled from the material, the difference may be too small to distinguish with the precision of the results.  It is also possible that the assumed PA is incorrect, in which case the pattern is most likely rotating more slowly than the material.

8.)  Other methods of solving the TW84 equations for shearing patterns were found to produce results in agreement with those obtained using the current method.

9.) Previous estimates that relied on the assumptions of the density-wave interpretation are inconsistent with the results obtained using the current method.

10.)  The results are inconsistent with fundamental assumptions in the density-wave interpretation  that are often used for finding spiral arm pattern speeds.  The results are more consistent with spiral structure theories that allow for shearing patterns.

11.)  The characteristic timescale for winding is $\sim$ 500 Myrs.  The winding dilemma is avoidable if the spiral pattern is a recurring transient feature.

\acknowledgements

The authors thank Gustaaf van Moorsel for providing the H{\hskip 1.5pt \footnotesize  I} data maps and the rotation curve for NGC 1365.  The authors also thank Se-Heon Oh for providing the rotation curves and the uncertainties of the orientation parameters for both Holmberg II and IC 2574.  Bryan Borchers, Oleg Maknin, and Rick Aster are acknowledged for insightful discussions about developing the new solution method, the statistics for the analysis, and interpreting the results.  The anonymous referee is acknowledged for comments that helped to improve this manuscript.  This research has made use of the National Radio Astronomy Observatory which is a facility of the National Science Foundation operated under cooperative agreement by Associated Universities, Inc; and the NASA/IPAC Extragalactic Database (NED) which is operated by the Jet Propulsion Laboratory, California Institute of Technology, under contract with the National Aeronautics and Space Administration.

\end{document}